\documentclass[aps,prl,twocolumn,superscriptaddress]{revtex4-2}

\usepackage{graphicx}
\usepackage{amsmath}
\usepackage{braket}
\usepackage[breaklinks]{hyperref}
\hypersetup{
colorlinks=true,
urlcolor=blue,
citecolor=blue,
linkcolor = blue
}

\newcommand{\supplementarysection}{%
  \setcounter{figure}{0}
  \renewcommand{\thefigure}{S\arabic{figure}}
  \section{Supplemental Material}
  \setcounter{table}{0}
  \renewcommand{\thetable}{S\Roman{table}}
  
}

\begin{document}


\title{Spectroscopy and Ground-State Transfer of Ultracold Bosonic $^{39}$K$^{133}$Cs Molecules}

  \author{Krzysztof P. Zamarski}%
  \thanks{These authors contributed equally to this work.}
 \author{Charly Beulenkamp}%
 \thanks{These authors contributed equally to this work.}
 \author{Yi Zeng}%
 \author{Manuele Landini}%
 \author{Hanns-Christoph N\"agerl}%
  \email{christoph.naegerl@uibk.ac.at}
\affiliation{Universit{\"a}t Innsbruck, Institut f{\"u}r Experimentalphysik und Zentrum f{\"u}r Quantenphysik, Technikerstraße 25, 6020 Innsbruck, Austria
}

\date{Jun 11, 2025}

\begin{abstract}
We report the creation of ultracold samples of $^{39}$K$^{133}$Cs molecules in their rovibrational ground state. By investigating potentially suitable excited states using one- and two-photon spectroscopy, we have identified a pathway to the ground state via an exceptionally narrow intermediate state. Using Stimulated Raman Adiabatic Passage (STIRAP), we create trapped samples of up to 3500 molecules at temperatures of 1 µK with one-way efficiencies of 71\%. The lifetime of these samples is limited by a near-universal two-body loss process, which could shed new light on similar loss mechanisms in other molecular species. Our results are a step towards establishing an alternative platform for the study of bosonic and fermionic quantum matter with strong dipolar interactions.
\end{abstract}

\maketitle

Ultracold dipolar molecules have recently emerged as a versatile experimental quantum physics platform \cite{Langen2024, Carroll2025, Picard2024, Liu2021}. Thanks to their long-range dipolar interactions and an accessible manifold of rotational states, which can be used as a pseudo-spin degree of freedom, they have found applications in the study of many-body dynamics \cite{Yan2013, Li2023, Christakis2023, Carroll2025} and show potential as a tool for quantum computation \cite{Gregory2024, Holland2023, Bao2023, Picard2024}. In addition, they can be used to study quantum chemistry \cite{Liu2021} and perform precision measurements of fundamental constants \cite{DeMille2024}.

Dipolar molecules can either be cooled directly \cite{Shuman2010, Hummon2013, Zhelyazkova2014, Lim2018, Wu2021, Jorapur2024}, or assembled from ultracold atoms \cite{Ni.2008, Aikawa2010, Takekoshi2014, Molony2014, Park2015, Guo2016, Rvachov2017, Voges2020, Cairncross2021, He2024}. The latter approach has achieved very high phase-space densities, leading to the formation of degenerate Fermi gases \cite{DeMarco2019, Duda2023, Cao2023} and Bose-Einstein condensates \cite{Bigagli2024}. In order to take full advantage of their properties, dipolar molecules must be prepared in their rovibrational ground state. Highly efficient transfer to deeply bound molecular states can be achieved using Stimulated Raman Adiabatic Passage (STIRAP) \cite{Danzl2008, Ni.2008}, which has by now been applied to several dipolar species \cite{Aikawa2010, Takekoshi2014, Molony2014, Park2015, Guo2016, Rvachov2017, Voges2020, Stevenson2023, He2024}.

Some of the referenced molecules \cite{Ni.2008, Aikawa2010, Rvachov2017, He2024} undergo exothermic chemical reactions, which intrinsically limit their lifetimes. The ones that do not, however, also exhibit near-universal two-body losses \cite{Takekoshi2014, Guo2016, Gregory2019, Gersema2021, Bause2021, Stevenson2023}. In some cases these have been proven to stem from photo-excitation of two-molecule complexes \cite{Gregory2020, Liu2020}, but the resulting models fail to predict the behavior of other systems \cite{Gersema2021, Bause2021}. Creating an ultracold sample of a new molecular species would shed more light on this problem \cite{Bause2023}. Regardless of their origin, it has been shown that these losses can be suppressed by applying a high DC electric field and confining the molecules to two dimensions \cite{Valtolina.2020} or tuning their interactions \cite{Li2021}, or by using circularly polarized microwave radiation \cite{Anderegg2021, Schindewolf2022, Lin2023, Bigagli2023}, all of which allow the molecules to be evaporatively cooled.

Among the chemically stable molecules, NaK has so far been the only available example with both fermionic and bosonic isotopologues \cite{Park2015, Voges2020}. Another such molecule would be KCs, which has a similar permanent dipole moment (1.9 Debye as opposed to 2.7 Debye \cite{Ladjimi2024}), but a much larger dynamic polarizability at 1064 nm \cite{Vexiau2017} and a much larger mass, allowing for the creation of deeper optical lattices with lower laser power. In addition, at 1037 nm the polarizabilities of the weakly bound and ground-state KCs molecules are predicted to be equal \cite{Vexiau2017}, which would facilitate the ground-state transfer.

In this Letter, we report on the production of bosonic $^{39}$K$^{133}$Cs molecules in their rovibrational ground state in a 1064-nm optical trap. Using STIRAP, we create up to 3500 molecules at temperatures around 1 µK, with one-way transfer efficiencies of up to 71\%. The intermediate state that we use for this purpose has the narrowest natural linewidth among all excited states used for STIRAP of bialkalis thus far, measured to be $2\pi \times 80(6)$ kHz. The polarizability ratio of the ground-state and weakly bound KCs molecules at 1064 nm is 0.93(4), which prevents the transfer from exciting excessive sample oscillations.

The experiments presented here start with an ultracold sample of weakly bound Feshbach molecules \cite{Charly}. They are created from a mixture of $^{39}$K and $^{133}$Cs atoms in their hyperfine ground states in a 1064-nm single-beam optical dipole trap. We first sweep the magnetic offset field, $\vec{B}$, oriented in the vertical direction, over a broad interspecies Feshbach resonance at $|\vec{B}|$ = 361.6 G. This causes a fraction of the atom pairs to enter an s-wave bound state with a total angular momentum projection of $M_F = 4$. Once the resonance is crossed, the remaining free atoms are removed from the trap by a force due to a magnetic-field gradient \cite{supp}. After the gradient is switched off, we wait until the field settles around $|\vec{B}|$ = 343 G, where the bound state is closed-channel dominated, and where the spectroscopy and STIRAP are performed. At the end of the sequence, $|\vec{B}|$ is swept back up across the resonance to break up the Feshbach molecules and allow the constituent atoms to be detected via absorption imaging.

\begin{figure}
\includegraphics[width=0.48\textwidth,keepaspectratio]{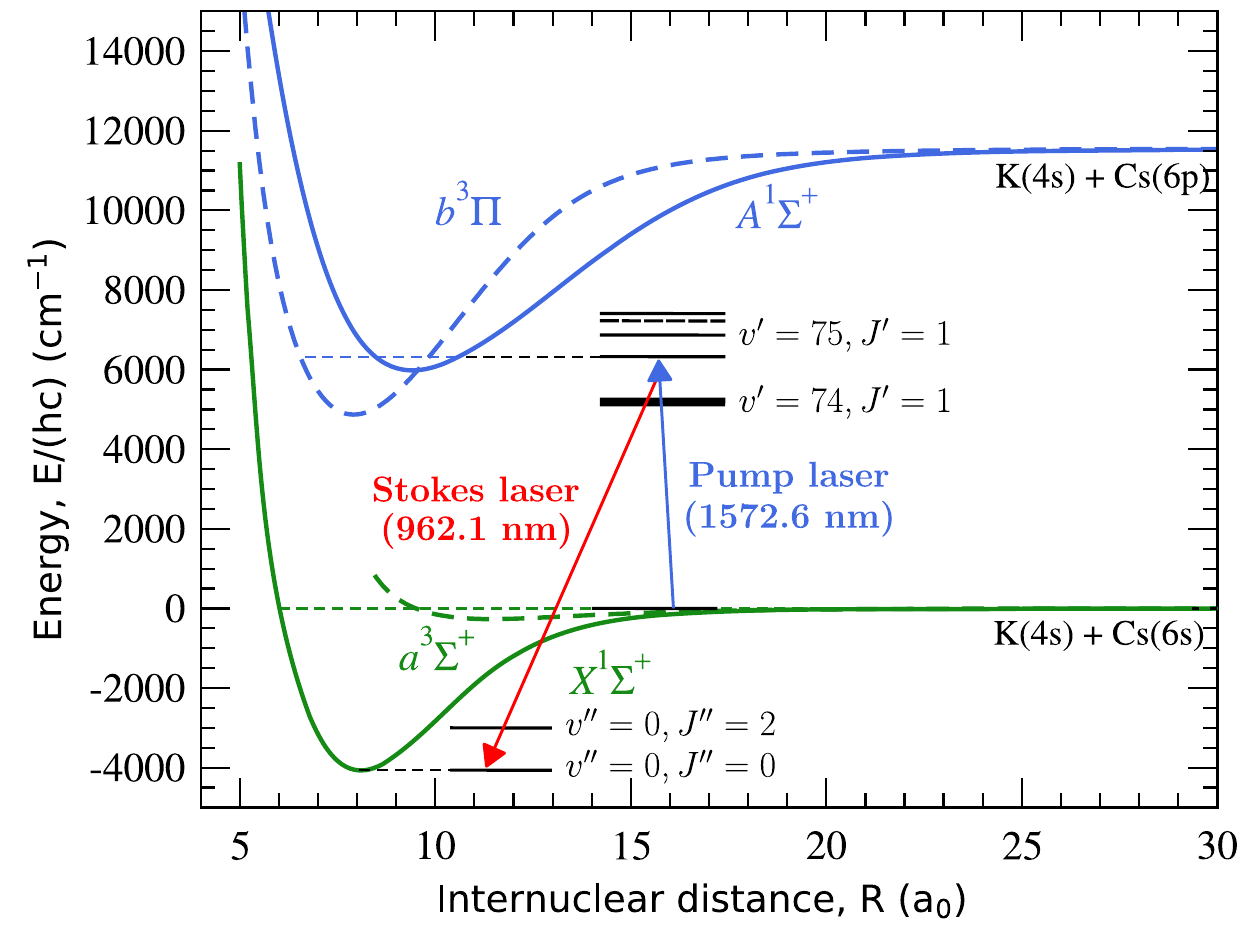}%
\caption{\label{fig1} Calculated potential energy curves of the $^{39}$K$^{133}$Cs molecule. Black lines show the levels investigated in this work (splittings within each potential are not drawn to scale), and the arrows represent the transitions that we use for STIRAP to the rovibrational ground state. Adapted from \cite{Borsalino2016} with permission.}
\end{figure}


\begin{figure*}
\includegraphics[width=\textwidth,keepaspectratio]{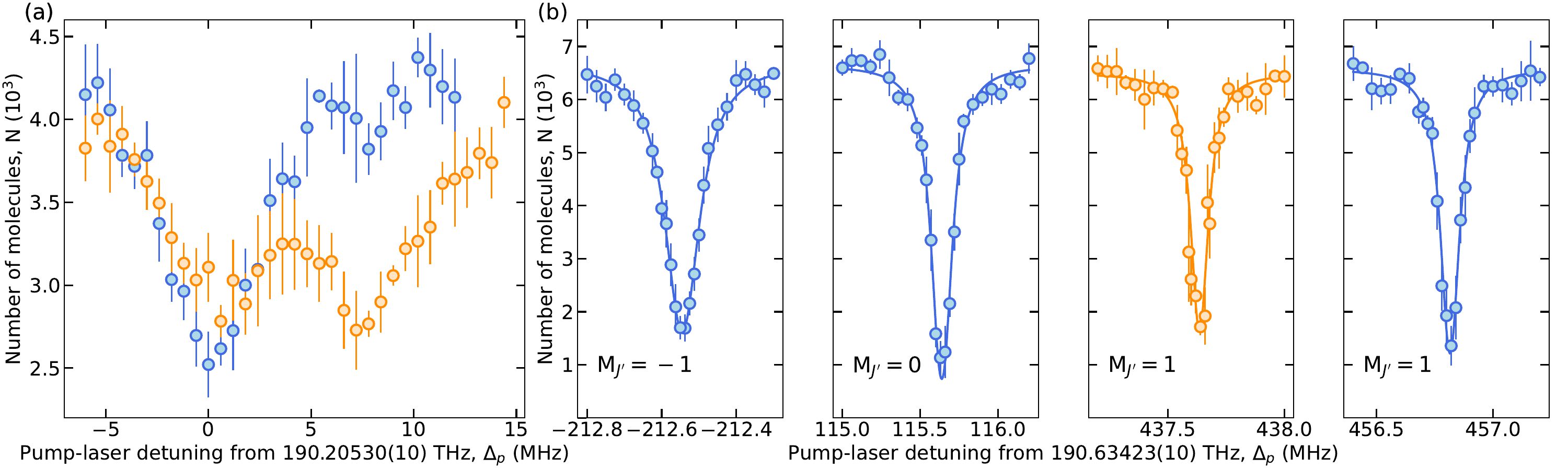}%
\caption{\label{fig2} Loss spectroscopy of the $^{39}$K$^{133}$Cs excited states. All measurements are performed at $|\vec{B}|$ = 343.03(6) G by exposing the sample to the pump-laser beam for 100 µs, and the laser power is adjusted to achieve a similar on-resonance lifetime for each state. Blue (orange) points are taken with the beam polarized horizontally (vertically). The error bars reflect the standard deviation from five experimental runs. (a) $v'=74$ state. The shapes of the observed loss features suggest that they comprise multiple unresolved hyperfine components. (b) Hyperfine-resolved spectroscopy of the $v'=75$ state. The solid lines are fits to the analytical model discussed in the text, yielding the linewidths and Rabi frequencies presented in Table I.}
\end{figure*}

For STIRAP we use two external-cavity diode lasers, later referred to as pump and Stokes laser. They are stabilized with respect to two separate high-finesse optical cavities using the Pound-Drever-Hall scheme \cite{Drever1983}. The cavities have finesses of $2.18(2)\times 10^5$ and $2.22(4)\times 10^5$, respectively, and occupy the same spacer, made of ultra-low expansion glass (ULE), which reduces the relative phase noise. In order to increase the optical power, the pump laser seeds a fiber amplifier, and the Stokes laser seeds a muti-mode laser diode. The two beams enter the experimental chamber horizontally in a nearly co-propagating configuration, perpendicular to the quantization axis, which is set by $\vec{B}$. They are both linearly polarized, so depending on the polarization direction they can address either $\pi$ or $\sigma^+$ and $\sigma^-$  transitions. Their waists are 83(2) µm and 32(2) µm, respectively.

\begin{table}[b]
\caption{\label{tab:table1} Comparison of the predicted and measured properties of two vibrational states of the $A^1\Sigma_0-b^3\Pi_\Omega$ potential of the $^{39}$K$^{133}$Cs molecule \cite{Nadia}. Here, $v'$ is the vibrational quantum number, $A^1\Sigma_0, b^3\Pi_0$ and $b^3\Pi_1$ are the contributions of these three electronic states, TDM is the predicted transition dipole moment in units of $10^{-4} ea_0$, and E$_\textrm{th}$ and E$_\textrm{exp}$ are the predicted and measured transition energies from the initial Feshbach state at $|\vec{B}|$ = 343 G in units of $h \times \textrm{THz}$. The angular momentum $J'$ is equal to 1 for both states.}
\begin{ruledtabular}
\begin{tabular}{ccccccc}
\textrm{$v'$}&
\textrm{A$^1\Sigma_0$}&
\textrm{b$^3\Pi_0$}&
\textrm{b$^3\Pi_1$}&
\textrm{TDM}&
\textrm{E$_\textrm{th}$}&
\textrm{E$_\textrm{exp}$}\\
\colrule
74 & 0.824 & 0.176 & $<10^{-5}$ & 3.96 & 190.2053 & 190.20530(10) \\
75 & $<10^{-5}$ & $<10^{-5}$ & 1 & 4.92 & 190.6342 & 190.63423(10) \\
\end{tabular}
\end{ruledtabular}
\end{table}

\begin{table}[b]
\caption{\label{tab:table2} Properties of the observed hyperfine levels of the $A^1\Sigma_0-b^3\Pi_\Omega, v' = 75$ state. Here, p is the pump laser polarization (h stands for horizontal, v for vertical), $M_{J'}$ is the projection of the angular momentum $J'$, TDM$_\textrm{th}$ and TDM$_\textrm{exp}$ are the predicted and measured transition dipole moments in units of $10^{-4} ea_0$, $\Omega_0$ is the measured intensity-normalized Rabi frequency in units of $2\pi \times$ kHz $\times \sqrt{I/\textrm{mW cm}^{-2}}$, and $\Gamma$ is the measured natural linewidth in units of $2\pi \times$ kHz. The values of $\Omega_0$ measured for horizontal polarization have been multiplied by two to obtain the value expected for the optimal (circular) polarization.}
\begin{ruledtabular}
\begin{tabular}{cccccc}
\textrm{p}&
\textrm{$M_{J'}$}&
\textrm{TDM$_\textrm{th}$}&
\textrm{TDM$_\textrm{exp}$}&
\textrm{$\Omega_0$}&
\textrm{$\Gamma$}\\
\colrule
v & 1 & 4.92 & 2.8(2) & 0.31(2) & 65(6) \\
h & 1 & 4.92 & 4.3(4) & 0.48(4) & 65(8)\\
h & 0 & 4.92 & 3.0(2) & 0.34(3) & 96(9)\\
h & -1 & 4.92 & 5.8(4) & 0.65(4) & 80(6)\\
\end{tabular}
\end{ruledtabular}
\end{table}

Efficient STIRAP requires the identification of a suitable excited molecular state. This state needs to have sufficient overlap with both the Feshbach and ground states, long enough lifetime to allow adiabatic transfer, and sufficiently well-resolved hyperfine structure to avoid destructive interferences \cite{Vitanov1999, Guo2017}. The excited states of the KCs molecule have been characterized experimentally using heat-pipe spectroscopy \cite{Tamanis2010, Kruzins2010, Kruzins2013}, but so far direct molecular excitations have not been observed in the ultracold regime. Fig.\ \ref{fig1} shows the potential energy curves of KCs calculated based on the heat-pipe results \cite{Borsalino2016}. The lowest two potentials in the first electronically excited state, $A^1\Sigma_0$ and $b^3\Pi_\Omega$, are depicted in their bare form, but in reality they are strongly mixed by spin-orbit coupling, which is why transitions to these states are allowed both from triplet Feshbach states and the singlet ground state. In addition, the $b^3\Pi_\Omega$ potential consists of three sublevels ($b^3\Pi_0$, $b^3\Pi_1$ and $b^3\Pi_2$) with different values of $\Omega$, the projection of the total electronic angular momentum onto the molecular axis. If $\Omega=0$, the Zeeman effect is greatly reduced, and the hyperfine structure usually becomes difficult to resolve experimentally.

We have performed loss spectroscopy on samples of Feshbach molecules by illuminating them with the pump laser for 100 µs and measuring their remaining number (Fig.\ \ref{fig2}). The two groups of loss features that we have identified match the theoretically predicted parameters of two vibrational levels of the spin-orbit coupled $A^1\Sigma_0-b^3\Pi_\Omega$ potential presented in Table \ref{tab:table1} \cite{Nadia}. In both cases, we can only address the $J'=1$ angular momentum state ($'$ denotes an electronically excited state), since we are starting from an s-wave Feshbach molecule. As expected, due to its $\Omega=0$ character, we are unable to resolve any hyperfine sublevels of the $v'=74$ state (Fig.\ \hyperref[fig2]{2(a)}). This makes it unsuitable for STIRAP. As a consequence, we have focused on the $v'=75$ state, which has an almost pure $\Omega=1$ character.

Scanning the frequency of the pump laser in the vicinity of the observed $v'=75$ state transition has revealed four narrow loss features (Fig.\ \hyperref[fig2]{2(b)}). They are grouped in three energy regions, which can be assigned to the three different projections ($M_{J'}$) of the angular momentum $J'$. We fit each of these features to an analytical model of the form \cite{Debatin2011}
\begin{equation}
N(t) = N_0\textrm{exp}(-t\Omega_p^2\Gamma/(\Gamma^2+4\Delta_p^2)),
\end{equation}
where $N_0$ is the initial molecule number, $t$ is the laser exposure time, $\Omega_p$ is the Rabi frequency of the pump laser, $\Gamma$ is the natural linewidth of the excited state, and $\Delta_p$ is the detuning of the pump laser from resonance. From this fit, we extract the natural linewidth, $\Gamma$, of each level and calculate the corresponding Rabi frequencies using an independent measurement of the on-resonance lifetime \cite{supp}. Table \ref{tab:table2} summarizes the results of these calculations.

\begin{figure}
\includegraphics[width=0.48\textwidth,keepaspectratio]{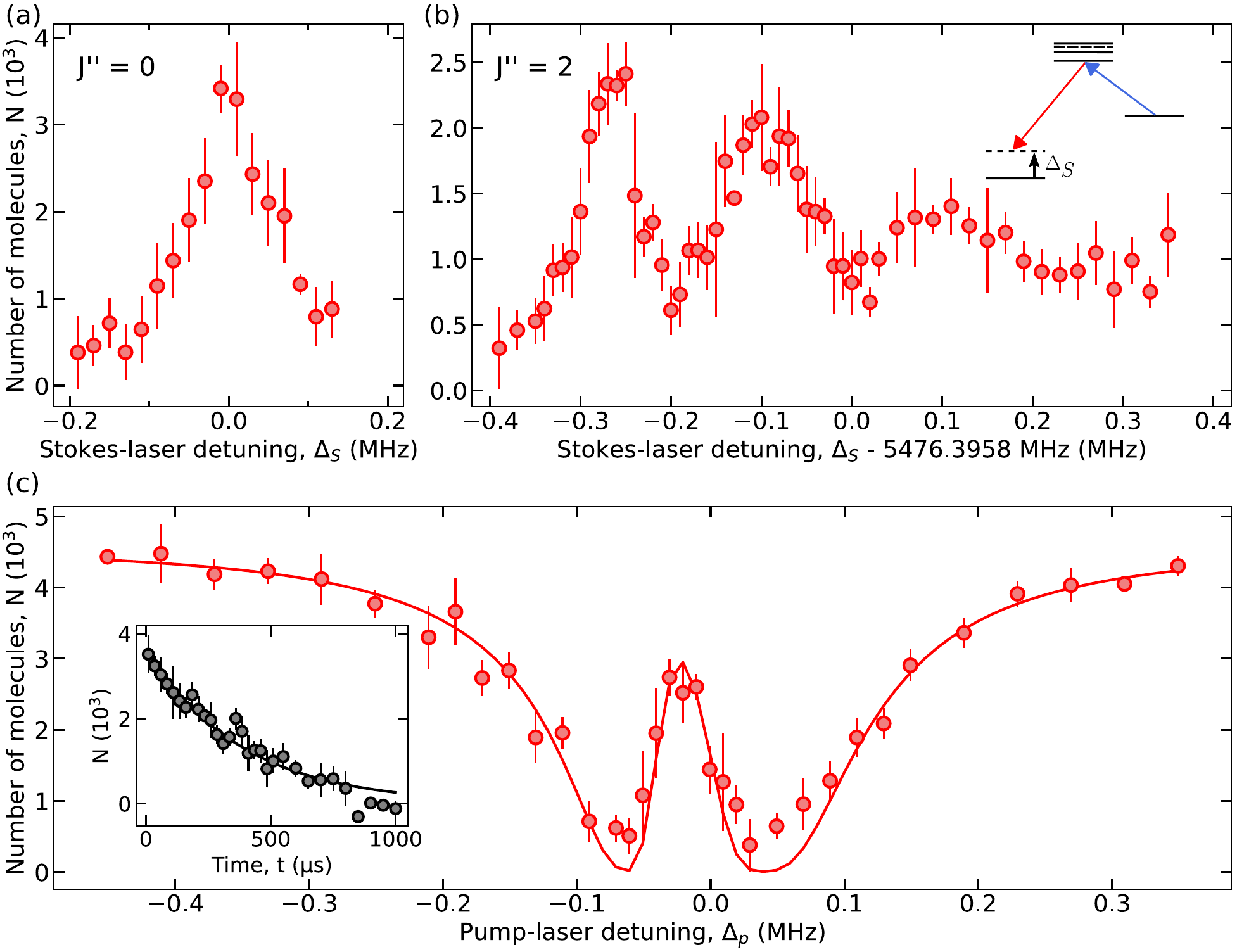}%
\caption{\label{fig3} Two-photon spectroscopy of the $^{39}$K$^{133}$Cs ground states. Each measurement is taken by exposing the sample to both laser beams for 100 µs, with the pump beam tuned to the $A^1\Sigma_0-b^3\Pi_\Omega$, $v'=75$, $M_{J'}=-1$ state. (a) Frequency scan of the Stokes beam, polarized horizontally, near the $v''=0, J''=0$ state. Zero detuning corresponds to 311.59709(5) THz. (b) Frequency scan of the Stokes beam, polarized vertically, near the $v''=0, J''=2$ state. We are able to resolve three unidentified hyperfine levels. (c) Dark resonance spectrum obtained by keeping the Stokes laser on resonance with the $v''=0, J''=0$ state and scanning the pump-laser frequency. The solid line is a fit on the basis of a numerical master equation model. Zero detuning is defined by the transition to the excited state and is extracted from the fit. The error bars reflect the standard deviation from five experimental runs. The inset shows the number of molecules as a function of laser exposure time on two-photon resonance, where the solid line is a fit based on the same model. The error bars reflect the standard deviation from three experimental runs.}
\end{figure}

A finite admixture of the $A^1\Sigma_0$ potential in the $v'=75$ state makes transitions to the $X^1\Sigma_0$, $v''=0$ manifold allowed ($''$ denotes a singlet electronic ground state). We are therefore able to measure the frequencies of these transitions by shining the Stokes laser while the pump laser is locked on resonance and observing a revival of the molecular population when the two-photon resonance is reached. Among the hyperfine sublevels of $v'=75$ that we have identified, the strongest coupling to $X^1\Sigma_0$, $v''=0$ is exhibited by the $M_{J'}=-1$ state, and the parity selection rule allows transitions from it to both the $J''=0$ and $J''=2$ levels in the ground state. We have identified two groups of resonances separated by 5.476 GHz (Fig.\ \hyperref[fig3]{3(a, b)}), which is consistent with the splitting expected between $J''=0$ and $J''=2$ according to the most recent calculation of the KCs rotational constant at equilibrium bond length of $B_e = 914.014(21)$ MHz \cite{Ferber2013}. Within the $J''=0$ state (Fig.\ \hyperref[fig3]{3(a)}), we are able to detect only one peak at 311.59709(5) THz, using a horizontally polarized beam. In $J''=2$ (Fig.\ \hyperref[fig3]{3(b)}), which we investigated with the beam polarized vertically, we are able to resolve three hyperfine levels, which is not enough to unambiguously assign their quantum numbers. The ground-state rotational constant extracted from this measurement is $B_0$ = $2\pi \times 912.68(18)$ MHz, which is reasonably close to the predicted $B_e$. The uncertainty is limited by our lack of knowledge of the hyperfine quantum numbers \cite{supp}.

In order to extract the strength of the detected transition to $J''=0$, we have performed an electromagnetically induced transparency (EIT \cite{Fleischhauer2005}) measurement by keeping the Stokes laser locked on resonance and scanning the frequency of the pump laser (Fig.\ \hyperref[fig3]{3(c)}). We fit the data to a numerical master equation model that takes into account the excited state linewidth, as well as a finite decoherence rate between the initial and final states \cite{supp}. The fit yields a normalized Rabi frequency of $2\pi \times 0.52(5)$ kHz $\times \sqrt{I/\textrm{mW cm}^{-2}}$ (assuming the optimal, circular polarization), which gives a transition dipole moment of 4.7(4) $10^{-4} ea_0$ (the predicted value is 4.75 $10^{-4} ea_0$ \cite{Nadia}). From a parallel measurement of the lifetime of molecules on the two-photon resonance (inset in Fig.\ \hyperref[fig3]{3(c)}) we extract an effective decoherence rate of $2\pi \times 6.3(4)$ kHz.

\begin{figure}
\includegraphics[width=0.48\textwidth,keepaspectratio]{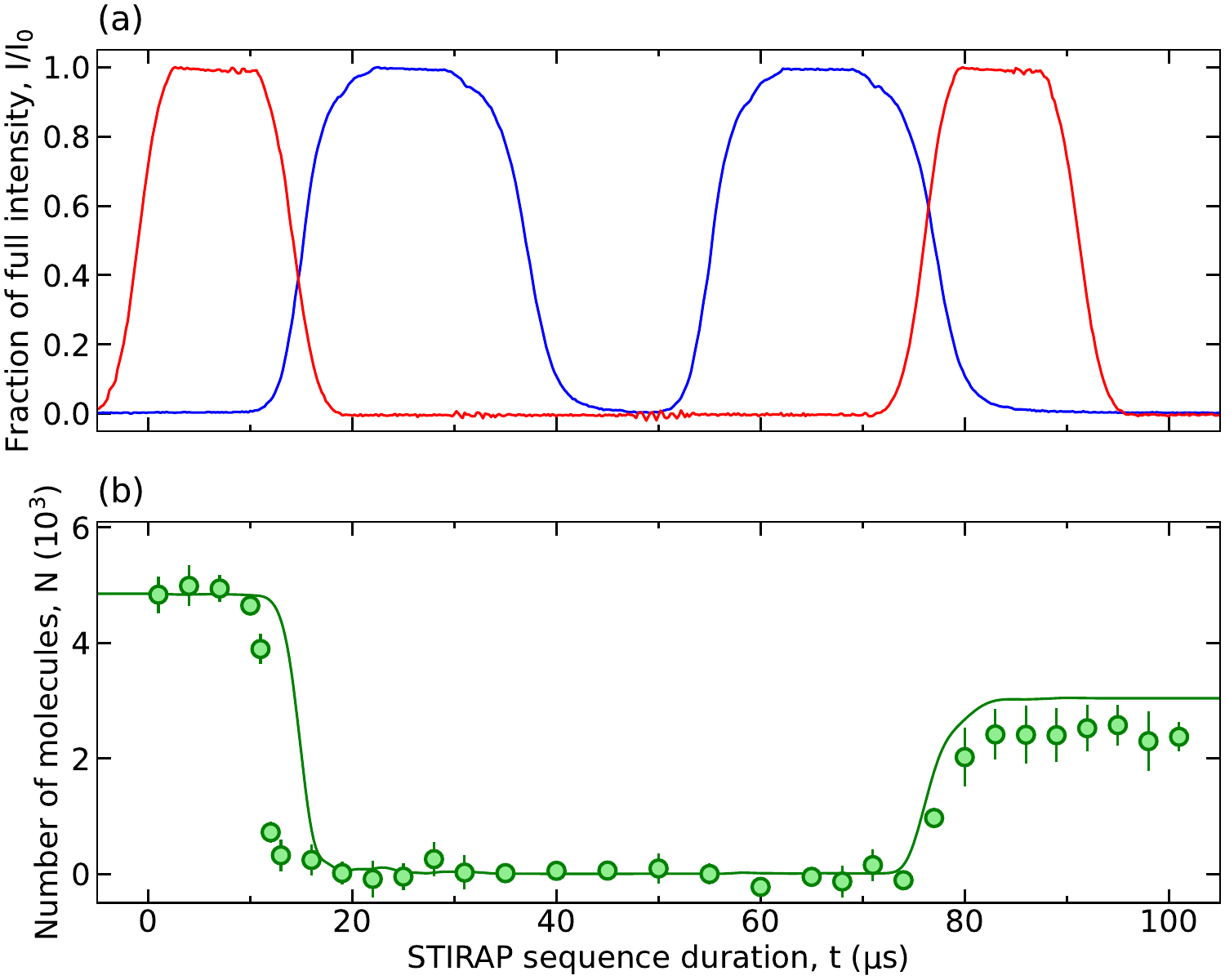}%
\caption{\label{fig4} STIRAP sequence. (a) Intensities of the pump beam (blue) and the Stokes beam (red) as a function of time, recorded by photodiodes and averaged over nine experimental runs. (b) Number of Feshbach molecules observed at different points of the pulse sequence by abruptly switching both beams off. The solid line is the result of a master equation simulation, in which the only free parameter is the initial number of molecules (see text for details). The error bars reflect the standard deviation from five experimental runs.}
\end{figure}

Having identified the necessary transitions, we proceed to transfer the molecules to the rovibrational ground state using STIRAP. The pulse sequence used (Fig.\ \hyperref[fig4]{4(a)}), usually referred to as ‘counterintuitive’, requires the Stokes laser to be switched on before the pump laser. The transfer to the ground state takes place as the pump laser power is being ramped up while the Stokes laser power is being ramped down, which causes the state of the molecules to adiabatically change character from the Feshbach state to the ground state. This process takes around 10 µs and is then reversed to allow the detection of remaining molecules. Fig.\ \hyperref[fig4]{4(b)} shows the evolution of the observed number of molecules as a typical STIRAP pulse sequence proceeds. The peak Rabi frequencies used were $2\pi \times 342(19)$ kHz for the pump beam and $2\pi \times 307(16)$ kHz for the Stokes beam. The ratio of the final and initial numbers allows us to estimate the one-way transfer efficiency to be around 71\%. The typical temperature of the molecules is around 1 µK at this stage and we do not observe any significant heating caused by the STIRAP pulses.

To assess our understanding of the transfer efficiency, we perform a master-equation simulation using the same model as for the fit in Fig.\ \hyperref[fig3]{3(c)} (solid line in Fig.\ \hyperref[fig4]{4(b)}), in which the only free parameter is now the initial number of molecules. We assume that the decoherence rate is independent of the laser power, but we are aware that at high enough power the effect of broadband laser noise will become significant \cite{Yatsenko2014}. The small discrepancy between the simulation and the data suggests that this is indeed the case, or that there are other relevant factors that have not been taken into account.

\begin{figure}
\includegraphics[width=0.48\textwidth,keepaspectratio]{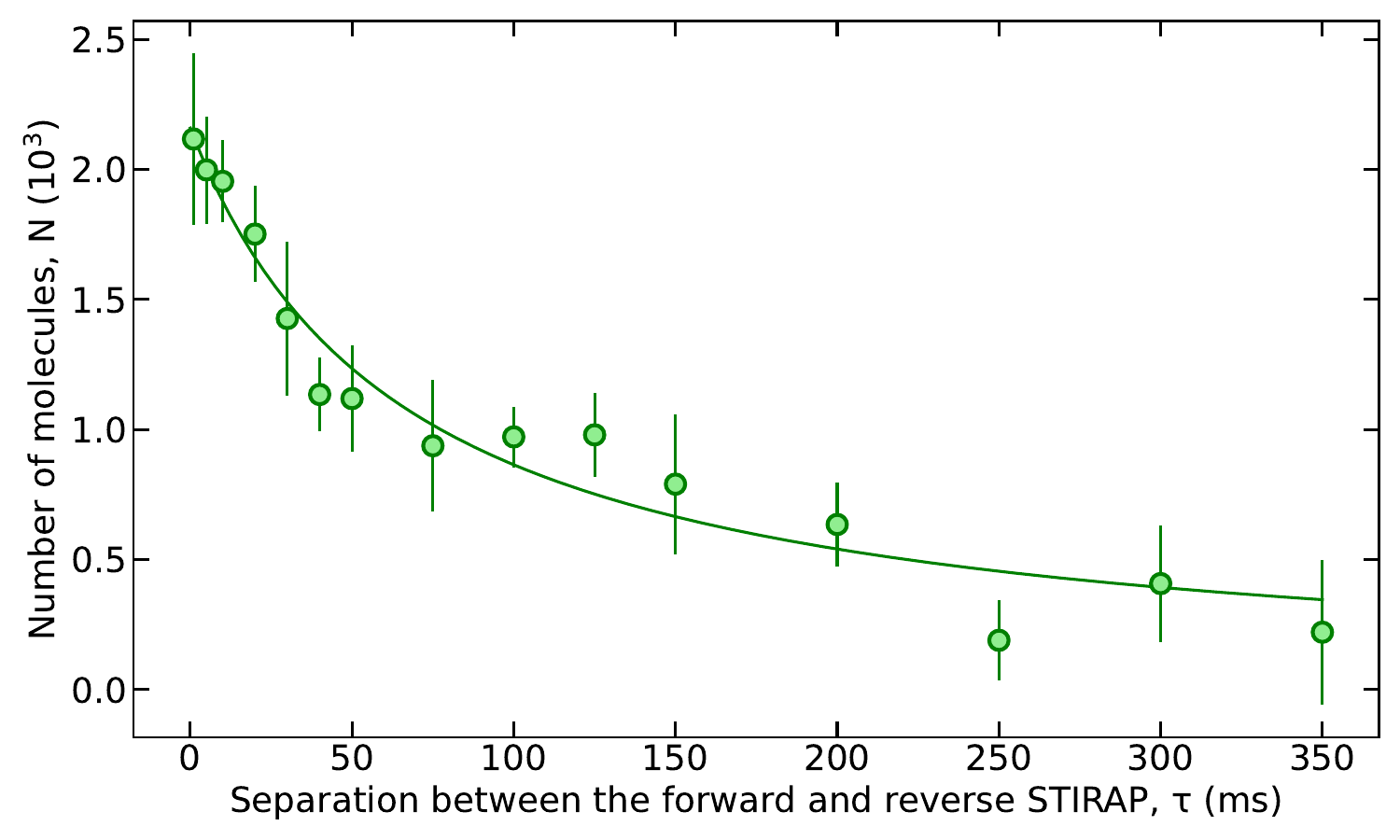}%
\caption{\label{fig5} Loss of $^{39}$K$^{133}$Cs molecules from a crossed optical trap. The initial peak density is $5.4(15)_{\textrm{sys}}(8)_{\textrm{stat}} \times 10^{10}$ cm$^{-3}$. The solid line is a fit to a two-body decay function, yielding a loss coefficient of $8(2)_{\textrm{sys}}(2)_{\textrm{stat}} \times 10^{-10}$ cm$^{3}$s$^{-1}$. The error bars reflect the standard deviation from five experimental runs.}
\end{figure}

One of the most important properties of a trapped sample of ground-state molecules is its lifetime. Before measuring it, in order to make sure that the molecules do not spread out along the optical trap, we add a perpendicular beam of a larger waist, which results in trap frequencies of $2\pi \times (10.6(1), 168(4), 168(4))$ Hz. We measure a polarizability of 78(11) Hz/(W/cm$^2$) at 1064 nm in the ground state, which is consistent with the prediction of 77.83 Hz/(W/cm$^2$) \cite{Vexiau2017}, and is only a factor of 0.93(4) smaller than for the Feshbach molecules \cite{supp}. The temperature under these conditions is 830(89) nK, giving a peak density of $5.4(15)_{\textrm{sys}}(8)_{\textrm{stat}} \times 10^{10}$ cm$^{-3}$. We then vary the separation, $\tau$, between the forward and reverse STIRAP, and measure the final number of molecules (Fig.\ \hyperref[fig5]{5}). Due to a significant drift of the magnetic field in our experimental chamber, we have to adjust the relative detuning of the STIRAP lasers for each pulse separation \cite{supp}. The results fit to a two-body decay curve with a loss coefficient of $8(2)_{\textrm{sys}}(2)_{\textrm{stat}} \times 10^{-10}$ cm$^{3}$s$^{-1}$ \cite{supp}, which is higher than the predicted universal loss rate of $2.8 \times 10^{-10}$ cm$^{3}$s$^{-1}$ \cite{Julienne2011}. Such a high value suggests that KCs could be undergoing the same photo-induced loss mechanism as, e.g., RbCs \cite{Gregory2020}. 

In conclusion, we have characterized two distinct excited states of the $^{39}$K$^{133}$Cs molecule and the transitions from one of them to the rovibrational ground state. Using STIRAP, we have prepared 3500 ground-state molecules at a temperature of 1 µK. Thanks to the exceptionally narrow linewidth of the excited state that we have used for this transfer ($2\pi \times 80(6)$ kHz), we have been able to achieve efficiencies up to 71\% with lower Rabi frequencies than in other experiments. The identified states are a good starting point for further spectroscopic study of KCs, which would allow a more accurate calculation of its hyperfine parameters. Similarly to other chemically stable molecules, KCs undergoes two-body collisional losses, so it could provide new insights into the details of this process. In the future, we are planning to test existing shielding techniques \cite{Valtolina.2020, Li2021, Bigagli2023} on ultracold KCs samples, which would make them a new tool for the study of dipolar quantum matter.

\begin{acknowledgments}
 We are grateful to Nadia Bouloufa-Maafa and Romain Vexiau for extending their excited state calculations to energies accessible in our experiment. We would also like to thank Govind Unnikrishnan, Dechao Zhang, Emil Kirilov and Philipp Schindler for technical contributions to the experiment and Andreas Schindewolf for critical reading of the manuscript.
 We acknowledge funding from the European Research Council (ERC) under Project No.\ 789017, the FWF under Project No.\ P29602-N36, a Wittgenstein prize grant under FWF Project No.\ Z336-N36, an FFG infrastructure grant with project number FO999896041 and by the FWF’s COE 1 and quantA. K.P.Z. acknowledges support from the Austrian Science Fund (FWF) within the DK-ALM (Grant No.\ W1259-N27).
This research was funded in whole or in part by the Austrian Science Fund (FWF) Grant DOI 10.55776/W1259. For open access purposes, the authors have applied a CC BY public copyright license to any author-accepted manuscript version arising from this submission.

Data supporting this study is publicly available from Zenodo at \href{https://zenodo.org/records/15638904}{10.5281/zenodo.15638904}.
\end{acknowledgments}

\bibliography{Refs}

\begin{thebibliography}{65}%
\makeatletter
\providecommand \@ifxundefined [1]{%
 \@ifx{#1\undefined}
}%
\providecommand \@ifnum [1]{%
 \ifnum #1\expandafter \@firstoftwo
 \else \expandafter \@secondoftwo
 \fi
}%
\providecommand \@ifx [1]{%
 \ifx #1\expandafter \@firstoftwo
 \else \expandafter \@secondoftwo
 \fi
}%
\providecommand \natexlab [1]{#1}%
\providecommand \enquote  [1]{``#1''}%
\providecommand \bibnamefont  [1]{#1}%
\providecommand \bibfnamefont [1]{#1}%
\providecommand \citenamefont [1]{#1}%
\providecommand \href@noop [0]{\@secondoftwo}%
\providecommand \href [0]{\begingroup \@sanitize@url \@href}%
\providecommand \@href[1]{\@@startlink{#1}\@@href}%
\providecommand \@@href[1]{\endgroup#1\@@endlink}%
\providecommand \@sanitize@url [0]{\catcode `\\12\catcode `\$12\catcode `\&12\catcode `\#12\catcode `\^12\catcode `\_12\catcode `\%12\relax}%
\providecommand \@@startlink[1]{}%
\providecommand \@@endlink[0]{}%
\providecommand \url  [0]{\begingroup\@sanitize@url \@url }%
\providecommand \@url [1]{\endgroup\@href {#1}{\urlprefix }}%
\providecommand \urlprefix  [0]{URL }%
\providecommand \Eprint [0]{\href }%
\providecommand \doibase [0]{https://doi.org/}%
\providecommand \selectlanguage [0]{\@gobble}%
\providecommand \bibinfo  [0]{\@secondoftwo}%
\providecommand \bibfield  [0]{\@secondoftwo}%
\providecommand \translation [1]{[#1]}%
\providecommand \BibitemOpen [0]{}%
\providecommand \bibitemStop [0]{}%
\providecommand \bibitemNoStop [0]{.\EOS\space}%
\providecommand \EOS [0]{\spacefactor3000\relax}%
\providecommand \BibitemShut  [1]{\csname bibitem#1\endcsname}%
\let\auto@bib@innerbib\@empty
\bibitem [{\citenamefont {Langen}\ \emph {et~al.}(2024)\citenamefont {Langen}, \citenamefont {Valtolina}, \citenamefont {Wang},\ and\ \citenamefont {Ye}}]{Langen2024}%
  \BibitemOpen
  \bibfield  {author} {\bibinfo {author} {\bibfnamefont {T.}~\bibnamefont {Langen}}, \bibinfo {author} {\bibfnamefont {G.}~\bibnamefont {Valtolina}}, \bibinfo {author} {\bibfnamefont {D.}~\bibnamefont {Wang}},\ and\ \bibinfo {author} {\bibfnamefont {J.}~\bibnamefont {Ye}},\ }\bibfield  {title} {\bibinfo {title} {{Quantum state manipulation and cooling of ultracold molecules}},\ }\href {https://doi.org/10.1038/s41567-024-02423-1} {\bibfield  {journal} {\bibinfo  {journal} {Nature Physics}\ }\textbf {\bibinfo {volume} {20}},\ \bibinfo {pages} {702} (\bibinfo {year} {2024})}\BibitemShut {NoStop}%
\bibitem [{\citenamefont {Carroll}\ \emph {et~al.}(2025)\citenamefont {Carroll}, \citenamefont {Hirzler}, \citenamefont {Miller}, \citenamefont {Wellnitz}, \citenamefont {Muleady}, \citenamefont {Lin}, \citenamefont {Zamarski}, \citenamefont {Wang}, \citenamefont {Bohn}, \citenamefont {Rey},\ and\ \citenamefont {Ye}}]{Carroll2025}%
  \BibitemOpen
  \bibfield  {author} {\bibinfo {author} {\bibfnamefont {A.~N.}\ \bibnamefont {Carroll}}, \bibinfo {author} {\bibfnamefont {H.}~\bibnamefont {Hirzler}}, \bibinfo {author} {\bibfnamefont {C.}~\bibnamefont {Miller}}, \bibinfo {author} {\bibfnamefont {D.}~\bibnamefont {Wellnitz}}, \bibinfo {author} {\bibfnamefont {S.~R.}\ \bibnamefont {Muleady}}, \bibinfo {author} {\bibfnamefont {J.}~\bibnamefont {Lin}}, \bibinfo {author} {\bibfnamefont {K.~P.}\ \bibnamefont {Zamarski}}, \bibinfo {author} {\bibfnamefont {R.~R.~W.}\ \bibnamefont {Wang}}, \bibinfo {author} {\bibfnamefont {J.~L.}\ \bibnamefont {Bohn}}, \bibinfo {author} {\bibfnamefont {A.~M.}\ \bibnamefont {Rey}},\ and\ \bibinfo {author} {\bibfnamefont {J.}~\bibnamefont {Ye}},\ }\bibfield  {title} {\bibinfo {title} {{Observation of generalized t-J spin dynamics with tunable dipolar interactions}},\ }\href {https://doi.org/10.1126/science.adq0911} {\bibfield  {journal} {\bibinfo  {journal} {Science}\ }\textbf {\bibinfo {volume} {388}},\ \bibinfo {pages} {381}
  (\bibinfo {year} {2025})}\BibitemShut {NoStop}%
\bibitem [{\citenamefont {Picard}\ \emph {et~al.}(2025)\citenamefont {Picard}, \citenamefont {Park}, \citenamefont {Patenotte}, \citenamefont {Gebretsadkan}, \citenamefont {Wellnitz}, \citenamefont {Rey},\ and\ \citenamefont {Ni}}]{Picard2024}%
  \BibitemOpen
  \bibfield  {author} {\bibinfo {author} {\bibfnamefont {L.~R.~B.}\ \bibnamefont {Picard}}, \bibinfo {author} {\bibfnamefont {A.~J.}\ \bibnamefont {Park}}, \bibinfo {author} {\bibfnamefont {G.~E.}\ \bibnamefont {Patenotte}}, \bibinfo {author} {\bibfnamefont {S.}~\bibnamefont {Gebretsadkan}}, \bibinfo {author} {\bibfnamefont {D.}~\bibnamefont {Wellnitz}}, \bibinfo {author} {\bibfnamefont {A.~M.}\ \bibnamefont {Rey}},\ and\ \bibinfo {author} {\bibfnamefont {K.-K.}\ \bibnamefont {Ni}},\ }\bibfield  {title} {\bibinfo {title} {{Entanglement and iSWAP Gate between Molecular Qubits}},\ }\href {https://doi.org/10.1038/s41586-024-08177-3} {\bibfield  {journal} {\bibinfo  {journal} {Nature}\ }\textbf {\bibinfo {volume} {637}},\ \bibinfo {pages} {821} (\bibinfo {year} {2025})}\BibitemShut {NoStop}%
\bibitem [{\citenamefont {Liu}\ \emph {et~al.}(2021)\citenamefont {Liu}, \citenamefont {Hu}, \citenamefont {Nichols}, \citenamefont {Yang}, \citenamefont {Xie}, \citenamefont {Guo},\ and\ \citenamefont {Ni}}]{Liu2021}%
  \BibitemOpen
  \bibfield  {author} {\bibinfo {author} {\bibfnamefont {Y.}~\bibnamefont {Liu}}, \bibinfo {author} {\bibfnamefont {M.~G.}\ \bibnamefont {Hu}}, \bibinfo {author} {\bibfnamefont {M.~A.}\ \bibnamefont {Nichols}}, \bibinfo {author} {\bibfnamefont {D.}~\bibnamefont {Yang}}, \bibinfo {author} {\bibfnamefont {D.}~\bibnamefont {Xie}}, \bibinfo {author} {\bibfnamefont {H.}~\bibnamefont {Guo}},\ and\ \bibinfo {author} {\bibfnamefont {K.~K.}\ \bibnamefont {Ni}},\ }\bibfield  {title} {\bibinfo {title} {{Precision test of statistical dynamics with state-to-state ultracold chemistry}},\ }\href {https://doi.org/10.1038/s41586-021-03459-6} {\bibfield  {journal} {\bibinfo  {journal} {Nature}\ }\textbf {\bibinfo {volume} {593}},\ \bibinfo {pages} {379} (\bibinfo {year} {2021})}\BibitemShut {NoStop}%
\bibitem [{\citenamefont {Yan}\ \emph {et~al.}(2013)\citenamefont {Yan}, \citenamefont {Moses}, \citenamefont {Gadway}, \citenamefont {Covey}, \citenamefont {Hazzard}, \citenamefont {Rey}, \citenamefont {Jin},\ and\ \citenamefont {Ye}}]{Yan2013}%
  \BibitemOpen
  \bibfield  {author} {\bibinfo {author} {\bibfnamefont {B.}~\bibnamefont {Yan}}, \bibinfo {author} {\bibfnamefont {S.~A.}\ \bibnamefont {Moses}}, \bibinfo {author} {\bibfnamefont {B.}~\bibnamefont {Gadway}}, \bibinfo {author} {\bibfnamefont {J.~P.}\ \bibnamefont {Covey}}, \bibinfo {author} {\bibfnamefont {K.~R.}\ \bibnamefont {Hazzard}}, \bibinfo {author} {\bibfnamefont {A.~M.}\ \bibnamefont {Rey}}, \bibinfo {author} {\bibfnamefont {D.~S.}\ \bibnamefont {Jin}},\ and\ \bibinfo {author} {\bibfnamefont {J.}~\bibnamefont {Ye}},\ }\bibfield  {title} {\bibinfo {title} {{Observation of dipolar spin-exchange interactions with lattice-confined polar molecules}},\ }\href {https://doi.org/10.1038/nature12483} {\bibfield  {journal} {\bibinfo  {journal} {Nature}\ }\textbf {\bibinfo {volume} {501}},\ \bibinfo {pages} {521} (\bibinfo {year} {2013})}\BibitemShut {NoStop}%
\bibitem [{\citenamefont {Li}\ \emph {et~al.}(2023)\citenamefont {Li}, \citenamefont {Matsuda}, \citenamefont {Miller}, \citenamefont {Carroll}, \citenamefont {Tobias}, \citenamefont {Higgins},\ and\ \citenamefont {Ye}}]{Li2023}%
  \BibitemOpen
  \bibfield  {author} {\bibinfo {author} {\bibfnamefont {J.-R.}\ \bibnamefont {Li}}, \bibinfo {author} {\bibfnamefont {K.}~\bibnamefont {Matsuda}}, \bibinfo {author} {\bibfnamefont {C.}~\bibnamefont {Miller}}, \bibinfo {author} {\bibfnamefont {A.~N.}\ \bibnamefont {Carroll}}, \bibinfo {author} {\bibfnamefont {W.~G.}\ \bibnamefont {Tobias}}, \bibinfo {author} {\bibfnamefont {J.~S.}\ \bibnamefont {Higgins}},\ and\ \bibinfo {author} {\bibfnamefont {J.}~\bibnamefont {Ye}},\ }\bibfield  {title} {\bibinfo {title} {{Tunable itinerant spin dynamics with polar molecules}},\ }\href {https://doi.org/10.1038/s41586-022-05479-2} {\bibfield  {journal} {\bibinfo  {journal} {Nature}\ }\textbf {\bibinfo {volume} {614}},\ \bibinfo {pages} {70} (\bibinfo {year} {2023})}\BibitemShut {NoStop}%
\bibitem [{\citenamefont {Christakis}\ \emph {et~al.}(2023)\citenamefont {Christakis}, \citenamefont {Rosenberg}, \citenamefont {Raj}, \citenamefont {Chi}, \citenamefont {Morningstar}, \citenamefont {Huse}, \citenamefont {Yan},\ and\ \citenamefont {Bakr}}]{Christakis2023}%
  \BibitemOpen
  \bibfield  {author} {\bibinfo {author} {\bibfnamefont {L.}~\bibnamefont {Christakis}}, \bibinfo {author} {\bibfnamefont {J.~S.}\ \bibnamefont {Rosenberg}}, \bibinfo {author} {\bibfnamefont {R.}~\bibnamefont {Raj}}, \bibinfo {author} {\bibfnamefont {S.}~\bibnamefont {Chi}}, \bibinfo {author} {\bibfnamefont {A.}~\bibnamefont {Morningstar}}, \bibinfo {author} {\bibfnamefont {D.~A.}\ \bibnamefont {Huse}}, \bibinfo {author} {\bibfnamefont {Z.~Z.}\ \bibnamefont {Yan}},\ and\ \bibinfo {author} {\bibfnamefont {W.~S.}\ \bibnamefont {Bakr}},\ }\bibfield  {title} {\bibinfo {title} {{Probing site-resolved correlations in a spin system of ultracold molecules}},\ }\href {https://doi.org/10.1038/s41586-022-05558-4} {\bibfield  {journal} {\bibinfo  {journal} {Nature}\ }\textbf {\bibinfo {volume} {614}},\ \bibinfo {pages} {64} (\bibinfo {year} {2023})}\BibitemShut {NoStop}%
\bibitem [{\citenamefont {Gregory}\ \emph {et~al.}(2024)\citenamefont {Gregory}, \citenamefont {Fernley}, \citenamefont {Tao}, \citenamefont {Bromley}, \citenamefont {Stepp}, \citenamefont {Zhang}, \citenamefont {Kotochigova}, \citenamefont {Hazzard},\ and\ \citenamefont {Cornish}}]{Gregory2024}%
  \BibitemOpen
  \bibfield  {author} {\bibinfo {author} {\bibfnamefont {P.~D.}\ \bibnamefont {Gregory}}, \bibinfo {author} {\bibfnamefont {L.~M.}\ \bibnamefont {Fernley}}, \bibinfo {author} {\bibfnamefont {A.~L.}\ \bibnamefont {Tao}}, \bibinfo {author} {\bibfnamefont {S.~L.}\ \bibnamefont {Bromley}}, \bibinfo {author} {\bibfnamefont {J.}~\bibnamefont {Stepp}}, \bibinfo {author} {\bibfnamefont {Z.}~\bibnamefont {Zhang}}, \bibinfo {author} {\bibfnamefont {S.}~\bibnamefont {Kotochigova}}, \bibinfo {author} {\bibfnamefont {K.~R.}\ \bibnamefont {Hazzard}},\ and\ \bibinfo {author} {\bibfnamefont {S.~L.}\ \bibnamefont {Cornish}},\ }\bibfield  {title} {\bibinfo {title} {{Second-scale rotational coherence and dipolar interactions in a gas of ultracold polar molecules}},\ }\href {https://doi.org/10.1038/s41567-023-02328-5} {\bibfield  {journal} {\bibinfo  {journal} {Nature Physics}\ }\textbf {\bibinfo {volume} {20}},\ \bibinfo {pages} {415} (\bibinfo {year} {2024})}\BibitemShut {NoStop}%
\bibitem [{\citenamefont {Holland}\ \emph {et~al.}(2023)\citenamefont {Holland}, \citenamefont {Lu},\ and\ \citenamefont {Cheuk}}]{Holland2023}%
  \BibitemOpen
  \bibfield  {author} {\bibinfo {author} {\bibfnamefont {C.~M.}\ \bibnamefont {Holland}}, \bibinfo {author} {\bibfnamefont {Y.}~\bibnamefont {Lu}},\ and\ \bibinfo {author} {\bibfnamefont {L.~W.}\ \bibnamefont {Cheuk}},\ }\bibfield  {title} {\bibinfo {title} {{On-demand entanglement of molecules in a reconfigurable optical tweezer array}},\ }\href {https://doi.org/10.1126/science.adf4272} {\bibfield  {journal} {\bibinfo  {journal} {Science}\ }\textbf {\bibinfo {volume} {382}},\ \bibinfo {pages} {1143} (\bibinfo {year} {2023})}\BibitemShut {NoStop}%
\bibitem [{\citenamefont {Bao}\ \emph {et~al.}(2023)\citenamefont {Bao}, \citenamefont {Yu}, \citenamefont {Anderegg}, \citenamefont {Chae}, \citenamefont {Ketterle}, \citenamefont {Ni},\ and\ \citenamefont {Doyle}}]{Bao2023}%
  \BibitemOpen
  \bibfield  {author} {\bibinfo {author} {\bibfnamefont {Y.}~\bibnamefont {Bao}}, \bibinfo {author} {\bibfnamefont {S.~S.}\ \bibnamefont {Yu}}, \bibinfo {author} {\bibfnamefont {L.}~\bibnamefont {Anderegg}}, \bibinfo {author} {\bibfnamefont {E.}~\bibnamefont {Chae}}, \bibinfo {author} {\bibfnamefont {W.}~\bibnamefont {Ketterle}}, \bibinfo {author} {\bibfnamefont {K.~K.}\ \bibnamefont {Ni}},\ and\ \bibinfo {author} {\bibfnamefont {J.~M.}\ \bibnamefont {Doyle}},\ }\bibfield  {title} {\bibinfo {title} {{Dipolar spin-exchange and entanglement between molecules in an optical tweezer array}},\ }\href {https://doi.org/10.1126/science.adf8999} {\bibfield  {journal} {\bibinfo  {journal} {Science}\ }\textbf {\bibinfo {volume} {382}},\ \bibinfo {pages} {1138} (\bibinfo {year} {2023})}\BibitemShut {NoStop}%
\bibitem [{\citenamefont {DeMille}\ \emph {et~al.}(2024)\citenamefont {DeMille}, \citenamefont {Hutzler}, \citenamefont {Rey},\ and\ \citenamefont {Zelevinsky}}]{DeMille2024}%
  \BibitemOpen
  \bibfield  {author} {\bibinfo {author} {\bibfnamefont {D.}~\bibnamefont {DeMille}}, \bibinfo {author} {\bibfnamefont {N.~R.}\ \bibnamefont {Hutzler}}, \bibinfo {author} {\bibfnamefont {A.~M.}\ \bibnamefont {Rey}},\ and\ \bibinfo {author} {\bibfnamefont {T.}~\bibnamefont {Zelevinsky}},\ }\bibfield  {title} {\bibinfo {title} {{Quantum sensing and metrology for fundamental physics with molecules}},\ }\href {https://doi.org/10.1038/s41567-024-02499-9} {\bibfield  {journal} {\bibinfo  {journal} {Nature Physics}\ }\textbf {\bibinfo {volume} {20}},\ \bibinfo {pages} {741} (\bibinfo {year} {2024})}\BibitemShut {NoStop}%
\bibitem [{\citenamefont {Shuman}\ \emph {et~al.}(2010)\citenamefont {Shuman}, \citenamefont {Barry},\ and\ \citenamefont {DeMille}}]{Shuman2010}%
  \BibitemOpen
  \bibfield  {author} {\bibinfo {author} {\bibfnamefont {E.~S.}\ \bibnamefont {Shuman}}, \bibinfo {author} {\bibfnamefont {J.~F.}\ \bibnamefont {Barry}},\ and\ \bibinfo {author} {\bibfnamefont {D.}~\bibnamefont {DeMille}},\ }\bibfield  {title} {\bibinfo {title} {{Laser cooling of a diatomic molecule}},\ }\href {https://doi.org/10.1038/nature09443} {\bibfield  {journal} {\bibinfo  {journal} {Nature}\ }\textbf {\bibinfo {volume} {467}},\ \bibinfo {pages} {820} (\bibinfo {year} {2010})}\BibitemShut {NoStop}%
\bibitem [{\citenamefont {Hummon}\ \emph {et~al.}(2013)\citenamefont {Hummon}, \citenamefont {Yeo}, \citenamefont {Stuhl}, \citenamefont {Collopy}, \citenamefont {Xia},\ and\ \citenamefont {Ye}}]{Hummon2013}%
  \BibitemOpen
  \bibfield  {author} {\bibinfo {author} {\bibfnamefont {M.~T.}\ \bibnamefont {Hummon}}, \bibinfo {author} {\bibfnamefont {M.}~\bibnamefont {Yeo}}, \bibinfo {author} {\bibfnamefont {B.~K.}\ \bibnamefont {Stuhl}}, \bibinfo {author} {\bibfnamefont {A.~L.}\ \bibnamefont {Collopy}}, \bibinfo {author} {\bibfnamefont {Y.}~\bibnamefont {Xia}},\ and\ \bibinfo {author} {\bibfnamefont {J.}~\bibnamefont {Ye}},\ }\bibfield  {title} {\bibinfo {title} {{2D Magneto-Optical Trapping of Diatomic Molecules}},\ }\href {https://doi.org/10.1103/PhysRevLett.110.143001} {\bibfield  {journal} {\bibinfo  {journal} {Physical Review Letters}\ }\textbf {\bibinfo {volume} {110}},\ \bibinfo {pages} {143001} (\bibinfo {year} {2013})}\BibitemShut {NoStop}%
\bibitem [{\citenamefont {Zhelyazkova}\ \emph {et~al.}(2014)\citenamefont {Zhelyazkova}, \citenamefont {Cournol}, \citenamefont {Wall}, \citenamefont {Matsushima}, \citenamefont {Hudson}, \citenamefont {Hinds}, \citenamefont {Tarbutt},\ and\ \citenamefont {Sauer}}]{Zhelyazkova2014}%
  \BibitemOpen
  \bibfield  {author} {\bibinfo {author} {\bibfnamefont {V.}~\bibnamefont {Zhelyazkova}}, \bibinfo {author} {\bibfnamefont {A.}~\bibnamefont {Cournol}}, \bibinfo {author} {\bibfnamefont {T.~E.}\ \bibnamefont {Wall}}, \bibinfo {author} {\bibfnamefont {A.}~\bibnamefont {Matsushima}}, \bibinfo {author} {\bibfnamefont {J.~J.}\ \bibnamefont {Hudson}}, \bibinfo {author} {\bibfnamefont {E.~A.}\ \bibnamefont {Hinds}}, \bibinfo {author} {\bibfnamefont {M.~R.}\ \bibnamefont {Tarbutt}},\ and\ \bibinfo {author} {\bibfnamefont {B.~E.}\ \bibnamefont {Sauer}},\ }\bibfield  {title} {\bibinfo {title} {{Laser cooling and slowing of CaF molecules}},\ }\href {https://doi.org/10.1103/PhysRevA.89.053416} {\bibfield  {journal} {\bibinfo  {journal} {Physical Review A}\ }\textbf {\bibinfo {volume} {89}},\ \bibinfo {pages} {053416} (\bibinfo {year} {2014})}\BibitemShut {NoStop}%
\bibitem [{\citenamefont {Lim}\ \emph {et~al.}(2018)\citenamefont {Lim}, \citenamefont {Almond}, \citenamefont {Trigatzis}, \citenamefont {Devlin}, \citenamefont {Fitch}, \citenamefont {Sauer}, \citenamefont {Tarbutt},\ and\ \citenamefont {Hinds}}]{Lim2018}%
  \BibitemOpen
  \bibfield  {author} {\bibinfo {author} {\bibfnamefont {J.}~\bibnamefont {Lim}}, \bibinfo {author} {\bibfnamefont {J.~R.}\ \bibnamefont {Almond}}, \bibinfo {author} {\bibfnamefont {M.~A.}\ \bibnamefont {Trigatzis}}, \bibinfo {author} {\bibfnamefont {J.~A.}\ \bibnamefont {Devlin}}, \bibinfo {author} {\bibfnamefont {N.~J.}\ \bibnamefont {Fitch}}, \bibinfo {author} {\bibfnamefont {B.~E.}\ \bibnamefont {Sauer}}, \bibinfo {author} {\bibfnamefont {M.~R.}\ \bibnamefont {Tarbutt}},\ and\ \bibinfo {author} {\bibfnamefont {E.~A.}\ \bibnamefont {Hinds}},\ }\bibfield  {title} {\bibinfo {title} {{Laser Cooled YbF Molecules for Measuring the Electron's Electric Dipole Moment}},\ }\href {https://doi.org/10.1103/PhysRevLett.120.123201} {\bibfield  {journal} {\bibinfo  {journal} {Physical Review Letters}\ }\textbf {\bibinfo {volume} {120}},\ \bibinfo {pages} {123201} (\bibinfo {year} {2018})}\BibitemShut {NoStop}%
\bibitem [{\citenamefont {Wu}\ \emph {et~al.}(2021)\citenamefont {Wu}, \citenamefont {Burau}, \citenamefont {Mehling}, \citenamefont {Ye},\ and\ \citenamefont {Ding}}]{Wu2021}%
  \BibitemOpen
  \bibfield  {author} {\bibinfo {author} {\bibfnamefont {Y.}~\bibnamefont {Wu}}, \bibinfo {author} {\bibfnamefont {J.~J.}\ \bibnamefont {Burau}}, \bibinfo {author} {\bibfnamefont {K.}~\bibnamefont {Mehling}}, \bibinfo {author} {\bibfnamefont {J.}~\bibnamefont {Ye}},\ and\ \bibinfo {author} {\bibfnamefont {S.}~\bibnamefont {Ding}},\ }\bibfield  {title} {\bibinfo {title} {{High Phase-Space Density of Laser-Cooled Molecules in an Optical Lattice}},\ }\href {https://doi.org/10.1103/physrevlett.127.263201} {\bibfield  {journal} {\bibinfo  {journal} {Physical Review Letters}\ }\textbf {\bibinfo {volume} {127}},\ \bibinfo {pages} {263201} (\bibinfo {year} {2021})}\BibitemShut {NoStop}%
\bibitem [{\citenamefont {Jorapur}\ \emph {et~al.}(2024)\citenamefont {Jorapur}, \citenamefont {Langin}, \citenamefont {Wang}, \citenamefont {Zheng},\ and\ \citenamefont {DeMille}}]{Jorapur2024}%
  \BibitemOpen
  \bibfield  {author} {\bibinfo {author} {\bibfnamefont {V.}~\bibnamefont {Jorapur}}, \bibinfo {author} {\bibfnamefont {T.~K.}\ \bibnamefont {Langin}}, \bibinfo {author} {\bibfnamefont {Q.}~\bibnamefont {Wang}}, \bibinfo {author} {\bibfnamefont {G.}~\bibnamefont {Zheng}},\ and\ \bibinfo {author} {\bibfnamefont {D.}~\bibnamefont {DeMille}},\ }\bibfield  {title} {\bibinfo {title} {{High Density Loading and Collisional Loss of Laser-Cooled Molecules in an Optical Trap}},\ }\href {https://doi.org/10.1103/PhysRevLett.132.163403} {\bibfield  {journal} {\bibinfo  {journal} {Physical Review Letters}\ }\textbf {\bibinfo {volume} {132}},\ \bibinfo {pages} {163403} (\bibinfo {year} {2024})}\BibitemShut {NoStop}%
\bibitem [{\citenamefont {Ni}\ \emph {et~al.}(2008)\citenamefont {Ni}, \citenamefont {Ospelkaus}, \citenamefont {de~Miranda}, \citenamefont {Pe'er}, \citenamefont {Neyenhuis}, \citenamefont {Zirbel}, \citenamefont {Kotochigova}, \citenamefont {Julienne}, \citenamefont {Jin},\ and\ \citenamefont {Ye}}]{Ni.2008}%
  \BibitemOpen
  \bibfield  {author} {\bibinfo {author} {\bibfnamefont {K.-K.}\ \bibnamefont {Ni}}, \bibinfo {author} {\bibfnamefont {S.}~\bibnamefont {Ospelkaus}}, \bibinfo {author} {\bibfnamefont {M.~H.~G.}\ \bibnamefont {de~Miranda}}, \bibinfo {author} {\bibfnamefont {A.}~\bibnamefont {Pe'er}}, \bibinfo {author} {\bibfnamefont {B.}~\bibnamefont {Neyenhuis}}, \bibinfo {author} {\bibfnamefont {J.~J.}\ \bibnamefont {Zirbel}}, \bibinfo {author} {\bibfnamefont {S.}~\bibnamefont {Kotochigova}}, \bibinfo {author} {\bibfnamefont {P.~S.}\ \bibnamefont {Julienne}}, \bibinfo {author} {\bibfnamefont {D.~S.}\ \bibnamefont {Jin}},\ and\ \bibinfo {author} {\bibfnamefont {J.}~\bibnamefont {Ye}},\ }\bibfield  {title} {\bibinfo {title} {{A High Phase-Space-Density Gas of Polar Molecules}},\ }\href {https://doi.org/10.1126/science.1163861} {\bibfield  {journal} {\bibinfo  {journal} {Science}\ }\textbf {\bibinfo {volume} {322}},\ \bibinfo {pages} {231} (\bibinfo {year} {2008})}\BibitemShut {NoStop}%
\bibitem [{\citenamefont {Aikawa}\ \emph {et~al.}(2010)\citenamefont {Aikawa}, \citenamefont {Akamatsu}, \citenamefont {Hayashi}, \citenamefont {Oasa}, \citenamefont {Kobayashi}, \citenamefont {Naidon}, \citenamefont {Kishimoto}, \citenamefont {Ueda},\ and\ \citenamefont {Inouye}}]{Aikawa2010}%
  \BibitemOpen
  \bibfield  {author} {\bibinfo {author} {\bibfnamefont {K.}~\bibnamefont {Aikawa}}, \bibinfo {author} {\bibfnamefont {D.}~\bibnamefont {Akamatsu}}, \bibinfo {author} {\bibfnamefont {M.}~\bibnamefont {Hayashi}}, \bibinfo {author} {\bibfnamefont {K.}~\bibnamefont {Oasa}}, \bibinfo {author} {\bibfnamefont {J.}~\bibnamefont {Kobayashi}}, \bibinfo {author} {\bibfnamefont {P.}~\bibnamefont {Naidon}}, \bibinfo {author} {\bibfnamefont {T.}~\bibnamefont {Kishimoto}}, \bibinfo {author} {\bibfnamefont {M.}~\bibnamefont {Ueda}},\ and\ \bibinfo {author} {\bibfnamefont {S.}~\bibnamefont {Inouye}},\ }\bibfield  {title} {\bibinfo {title} {{Coherent Transfer of Photoassociated Molecules into the Rovibrational Ground State}},\ }\href {https://doi.org/10.1103/PhysRevLett.105.203001} {\bibfield  {journal} {\bibinfo  {journal} {Physical Review Letters}\ }\textbf {\bibinfo {volume} {105}},\ \bibinfo {pages} {203001} (\bibinfo {year} {2010})}\BibitemShut {NoStop}%
\bibitem [{\citenamefont {Takekoshi}\ \emph {et~al.}(2014)\citenamefont {Takekoshi}, \citenamefont {Reichs{\"{o}}llner}, \citenamefont {Schindewolf}, \citenamefont {Hutson}, \citenamefont {{Le Sueur}}, \citenamefont {Dulieu}, \citenamefont {Ferlaino}, \citenamefont {Grimm},\ and\ \citenamefont {N{\"{a}}gerl}}]{Takekoshi2014}%
  \BibitemOpen
  \bibfield  {author} {\bibinfo {author} {\bibfnamefont {T.}~\bibnamefont {Takekoshi}}, \bibinfo {author} {\bibfnamefont {L.}~\bibnamefont {Reichs{\"{o}}llner}}, \bibinfo {author} {\bibfnamefont {A.}~\bibnamefont {Schindewolf}}, \bibinfo {author} {\bibfnamefont {J.~M.}\ \bibnamefont {Hutson}}, \bibinfo {author} {\bibfnamefont {C.~R.}\ \bibnamefont {{Le Sueur}}}, \bibinfo {author} {\bibfnamefont {O.}~\bibnamefont {Dulieu}}, \bibinfo {author} {\bibfnamefont {F.}~\bibnamefont {Ferlaino}}, \bibinfo {author} {\bibfnamefont {R.}~\bibnamefont {Grimm}},\ and\ \bibinfo {author} {\bibfnamefont {H.~C.}\ \bibnamefont {N{\"{a}}gerl}},\ }\bibfield  {title} {\bibinfo {title} {{Ultracold Dense Samples of Dipolar RbCs Molecules in the Rovibrational and Hyperfine Ground State}},\ }\href {https://doi.org/10.1103/PhysRevLett.113.205301} {\bibfield  {journal} {\bibinfo  {journal} {Physical Review Letters}\ }\textbf {\bibinfo {volume} {113}},\ \bibinfo {pages} {205301} (\bibinfo {year} {2014})}\BibitemShut {NoStop}%
\bibitem [{\citenamefont {Molony}\ \emph {et~al.}(2014)\citenamefont {Molony}, \citenamefont {Gregory}, \citenamefont {Ji}, \citenamefont {Lu}, \citenamefont {K{\"{o}}ppinger}, \citenamefont {{Le Sueur}}, \citenamefont {Blackley}, \citenamefont {Hutson},\ and\ \citenamefont {Cornish}}]{Molony2014}%
  \BibitemOpen
  \bibfield  {author} {\bibinfo {author} {\bibfnamefont {P.~K.}\ \bibnamefont {Molony}}, \bibinfo {author} {\bibfnamefont {P.~D.}\ \bibnamefont {Gregory}}, \bibinfo {author} {\bibfnamefont {Z.}~\bibnamefont {Ji}}, \bibinfo {author} {\bibfnamefont {B.}~\bibnamefont {Lu}}, \bibinfo {author} {\bibfnamefont {M.~P.}\ \bibnamefont {K{\"{o}}ppinger}}, \bibinfo {author} {\bibfnamefont {C.~R.}\ \bibnamefont {{Le Sueur}}}, \bibinfo {author} {\bibfnamefont {C.~L.}\ \bibnamefont {Blackley}}, \bibinfo {author} {\bibfnamefont {J.~M.}\ \bibnamefont {Hutson}},\ and\ \bibinfo {author} {\bibfnamefont {S.~L.}\ \bibnamefont {Cornish}},\ }\bibfield  {title} {\bibinfo {title} {{Creation of Ultracold $^{87}$Rb$^{133}$Cs Molecules in the Rovibrational Ground State}},\ }\href {https://doi.org/10.1103/PhysRevLett.113.255301} {\bibfield  {journal} {\bibinfo  {journal} {Physical Review Letters}\ }\textbf {\bibinfo {volume} {113}},\ \bibinfo {pages} {255301} (\bibinfo {year} {2014})}\BibitemShut {NoStop}%
\bibitem [{\citenamefont {Park}\ \emph {et~al.}(2015)\citenamefont {Park}, \citenamefont {Will},\ and\ \citenamefont {Zwierlein}}]{Park2015}%
  \BibitemOpen
  \bibfield  {author} {\bibinfo {author} {\bibfnamefont {J.~W.}\ \bibnamefont {Park}}, \bibinfo {author} {\bibfnamefont {S.~A.}\ \bibnamefont {Will}},\ and\ \bibinfo {author} {\bibfnamefont {M.~W.}\ \bibnamefont {Zwierlein}},\ }\bibfield  {title} {\bibinfo {title} {{Ultracold Dipolar Gas of Fermionic $^{23}$Na$^{40}$K Molecules in Their Absolute Ground State}},\ }\href {https://doi.org/10.1103/PhysRevLett.114.205302} {\bibfield  {journal} {\bibinfo  {journal} {Physical Review Letters}\ }\textbf {\bibinfo {volume} {114}},\ \bibinfo {pages} {205302} (\bibinfo {year} {2015})}\BibitemShut {NoStop}%
\bibitem [{\citenamefont {Guo}\ \emph {et~al.}(2016)\citenamefont {Guo}, \citenamefont {Zhu}, \citenamefont {Lu}, \citenamefont {Ye}, \citenamefont {Wang}, \citenamefont {Vexiau}, \citenamefont {Bouloufa-Maafa}, \citenamefont {Qu{\'{e}}m{\'{e}}ner}, \citenamefont {Dulieu},\ and\ \citenamefont {Wang}}]{Guo2016}%
  \BibitemOpen
  \bibfield  {author} {\bibinfo {author} {\bibfnamefont {M.}~\bibnamefont {Guo}}, \bibinfo {author} {\bibfnamefont {B.}~\bibnamefont {Zhu}}, \bibinfo {author} {\bibfnamefont {B.}~\bibnamefont {Lu}}, \bibinfo {author} {\bibfnamefont {X.}~\bibnamefont {Ye}}, \bibinfo {author} {\bibfnamefont {F.}~\bibnamefont {Wang}}, \bibinfo {author} {\bibfnamefont {R.}~\bibnamefont {Vexiau}}, \bibinfo {author} {\bibfnamefont {N.}~\bibnamefont {Bouloufa-Maafa}}, \bibinfo {author} {\bibfnamefont {G.}~\bibnamefont {Qu{\'{e}}m{\'{e}}ner}}, \bibinfo {author} {\bibfnamefont {O.}~\bibnamefont {Dulieu}},\ and\ \bibinfo {author} {\bibfnamefont {D.}~\bibnamefont {Wang}},\ }\bibfield  {title} {\bibinfo {title} {{Creation of an Ultracold Gas of Ground-State Dipolar $^{23}$Na$^{87}$Rb Molecules}},\ }\href {https://doi.org/10.1103/PhysRevLett.116.205303} {\bibfield  {journal} {\bibinfo  {journal} {Physical Review Letters}\ }\textbf {\bibinfo {volume} {116}},\ \bibinfo {pages} {205303} (\bibinfo {year} {2016})}\BibitemShut {NoStop}%
\bibitem [{\citenamefont {Rvachov}\ \emph {et~al.}(2017)\citenamefont {Rvachov}, \citenamefont {Son}, \citenamefont {Sommer}, \citenamefont {Ebadi}, \citenamefont {Park}, \citenamefont {Zwierlein}, \citenamefont {Ketterle},\ and\ \citenamefont {Jamison}}]{Rvachov2017}%
  \BibitemOpen
  \bibfield  {author} {\bibinfo {author} {\bibfnamefont {T.~M.}\ \bibnamefont {Rvachov}}, \bibinfo {author} {\bibfnamefont {H.}~\bibnamefont {Son}}, \bibinfo {author} {\bibfnamefont {A.~T.}\ \bibnamefont {Sommer}}, \bibinfo {author} {\bibfnamefont {S.}~\bibnamefont {Ebadi}}, \bibinfo {author} {\bibfnamefont {J.~J.}\ \bibnamefont {Park}}, \bibinfo {author} {\bibfnamefont {M.~W.}\ \bibnamefont {Zwierlein}}, \bibinfo {author} {\bibfnamefont {W.}~\bibnamefont {Ketterle}},\ and\ \bibinfo {author} {\bibfnamefont {A.~O.}\ \bibnamefont {Jamison}},\ }\bibfield  {title} {\bibinfo {title} {{Long-Lived Ultracold Molecules with Electric and Magnetic Dipole Moments}},\ }\href {https://doi.org/10.1103/PhysRevLett.119.143001} {\bibfield  {journal} {\bibinfo  {journal} {Physical Review Letters}\ }\textbf {\bibinfo {volume} {119}},\ \bibinfo {pages} {143001} (\bibinfo {year} {2017})}\BibitemShut {NoStop}%
\bibitem [{\citenamefont {Voges}\ \emph {et~al.}(2020)\citenamefont {Voges}, \citenamefont {Gersema}, \citenamefont {{Meyer Zum Alten Borgloh}}, \citenamefont {Schulze}, \citenamefont {Hartmann}, \citenamefont {Zenesini},\ and\ \citenamefont {Ospelkaus}}]{Voges2020}%
  \BibitemOpen
  \bibfield  {author} {\bibinfo {author} {\bibfnamefont {K.~K.}\ \bibnamefont {Voges}}, \bibinfo {author} {\bibfnamefont {P.}~\bibnamefont {Gersema}}, \bibinfo {author} {\bibfnamefont {M.}~\bibnamefont {{Meyer Zum Alten Borgloh}}}, \bibinfo {author} {\bibfnamefont {T.~A.}\ \bibnamefont {Schulze}}, \bibinfo {author} {\bibfnamefont {T.}~\bibnamefont {Hartmann}}, \bibinfo {author} {\bibfnamefont {A.}~\bibnamefont {Zenesini}},\ and\ \bibinfo {author} {\bibfnamefont {S.}~\bibnamefont {Ospelkaus}},\ }\bibfield  {title} {\bibinfo {title} {{Ultracold Gas of Bosonic $^{23}$Na$^{39}$K Ground-State Molecules}},\ }\href {https://doi.org/10.1103/PhysRevLett.125.083401} {\bibfield  {journal} {\bibinfo  {journal} {Physical Review Letters}\ }\textbf {\bibinfo {volume} {125}},\ \bibinfo {pages} {83401} (\bibinfo {year} {2020})}\BibitemShut {NoStop}%
\bibitem [{\citenamefont {Cairncross}\ \emph {et~al.}(2021)\citenamefont {Cairncross}, \citenamefont {Zhang}, \citenamefont {Picard}, \citenamefont {Yu}, \citenamefont {Wang},\ and\ \citenamefont {Ni}}]{Cairncross2021}%
  \BibitemOpen
  \bibfield  {author} {\bibinfo {author} {\bibfnamefont {W.~B.}\ \bibnamefont {Cairncross}}, \bibinfo {author} {\bibfnamefont {J.~T.}\ \bibnamefont {Zhang}}, \bibinfo {author} {\bibfnamefont {L.~R.}\ \bibnamefont {Picard}}, \bibinfo {author} {\bibfnamefont {Y.}~\bibnamefont {Yu}}, \bibinfo {author} {\bibfnamefont {K.}~\bibnamefont {Wang}},\ and\ \bibinfo {author} {\bibfnamefont {K.~K.}\ \bibnamefont {Ni}},\ }\bibfield  {title} {\bibinfo {title} {{Assembly of a Rovibrational Ground State Molecule in an Optical Tweezer}},\ }\href {https://doi.org/10.1103/PhysRevLett.126.123402} {\bibfield  {journal} {\bibinfo  {journal} {Physical Review Letters}\ }\textbf {\bibinfo {volume} {126}},\ \bibinfo {pages} {123402} (\bibinfo {year} {2021})}\BibitemShut {NoStop}%
\bibitem [{\citenamefont {He}\ \emph {et~al.}(2024)\citenamefont {He}, \citenamefont {Nie}, \citenamefont {Avalos}, \citenamefont {Botsi}, \citenamefont {Kumar}, \citenamefont {Yang},\ and\ \citenamefont {Dieckmann}}]{He2024}%
  \BibitemOpen
  \bibfield  {author} {\bibinfo {author} {\bibfnamefont {C.}~\bibnamefont {He}}, \bibinfo {author} {\bibfnamefont {X.}~\bibnamefont {Nie}}, \bibinfo {author} {\bibfnamefont {V.}~\bibnamefont {Avalos}}, \bibinfo {author} {\bibfnamefont {S.}~\bibnamefont {Botsi}}, \bibinfo {author} {\bibfnamefont {S.}~\bibnamefont {Kumar}}, \bibinfo {author} {\bibfnamefont {A.}~\bibnamefont {Yang}},\ and\ \bibinfo {author} {\bibfnamefont {K.}~\bibnamefont {Dieckmann}},\ }\bibfield  {title} {\bibinfo {title} {{Efficient Creation of Ultracold Ground State $^{6}$Li$^{40}$K Polar Molecules}},\ }\href {https://doi.org/10.1103/PhysRevLett.132.243401} {\bibfield  {journal} {\bibinfo  {journal} {Physical Review Letters}\ }\textbf {\bibinfo {volume} {132}},\ \bibinfo {pages} {243401} (\bibinfo {year} {2024})}\BibitemShut {NoStop}%
\bibitem [{\citenamefont {{De Marco}}\ \emph {et~al.}(2019)\citenamefont {{De Marco}}, \citenamefont {Valtolina}, \citenamefont {Matsuda}, \citenamefont {Tobias}, \citenamefont {Covey},\ and\ \citenamefont {Ye}}]{DeMarco2019}%
  \BibitemOpen
  \bibfield  {author} {\bibinfo {author} {\bibfnamefont {L.}~\bibnamefont {{De Marco}}}, \bibinfo {author} {\bibfnamefont {G.}~\bibnamefont {Valtolina}}, \bibinfo {author} {\bibfnamefont {K.}~\bibnamefont {Matsuda}}, \bibinfo {author} {\bibfnamefont {W.~G.}\ \bibnamefont {Tobias}}, \bibinfo {author} {\bibfnamefont {J.~P.}\ \bibnamefont {Covey}},\ and\ \bibinfo {author} {\bibfnamefont {J.}~\bibnamefont {Ye}},\ }\bibfield  {title} {\bibinfo {title} {{A degenerate Fermi gas of polar molecules}},\ }\href {https://doi.org/10.1126/science.aau7230} {\bibfield  {journal} {\bibinfo  {journal} {Science}\ }\textbf {\bibinfo {volume} {363}},\ \bibinfo {pages} {853} (\bibinfo {year} {2019})}\BibitemShut {NoStop}%
\bibitem [{\citenamefont {Duda}\ \emph {et~al.}(2023)\citenamefont {Duda}, \citenamefont {Chen}, \citenamefont {Schindewolf}, \citenamefont {Bause}, \citenamefont {von Milczewski}, \citenamefont {Schmidt}, \citenamefont {Bloch},\ and\ \citenamefont {Luo}}]{Duda2023}%
  \BibitemOpen
  \bibfield  {author} {\bibinfo {author} {\bibfnamefont {M.}~\bibnamefont {Duda}}, \bibinfo {author} {\bibfnamefont {X.~Y.}\ \bibnamefont {Chen}}, \bibinfo {author} {\bibfnamefont {A.}~\bibnamefont {Schindewolf}}, \bibinfo {author} {\bibfnamefont {R.}~\bibnamefont {Bause}}, \bibinfo {author} {\bibfnamefont {J.}~\bibnamefont {von Milczewski}}, \bibinfo {author} {\bibfnamefont {R.}~\bibnamefont {Schmidt}}, \bibinfo {author} {\bibfnamefont {I.}~\bibnamefont {Bloch}},\ and\ \bibinfo {author} {\bibfnamefont {X.~Y.}\ \bibnamefont {Luo}},\ }\bibfield  {title} {\bibinfo {title} {{Transition from a polaronic condensate to a degenerate Fermi gas of heteronuclear molecules}},\ }\href {https://doi.org/10.1038/s41567-023-01948-1} {\bibfield  {journal} {\bibinfo  {journal} {Nature Physics}\ }\textbf {\bibinfo {volume} {19}},\ \bibinfo {pages} {720} (\bibinfo {year} {2023})}\BibitemShut {NoStop}%
\bibitem [{\citenamefont {Cao}\ \emph {et~al.}(2023)\citenamefont {Cao}, \citenamefont {Yang}, \citenamefont {Su}, \citenamefont {Wang}, \citenamefont {Rui}, \citenamefont {Zhao},\ and\ \citenamefont {Pan}}]{Cao2023}%
  \BibitemOpen
  \bibfield  {author} {\bibinfo {author} {\bibfnamefont {J.}~\bibnamefont {Cao}}, \bibinfo {author} {\bibfnamefont {H.}~\bibnamefont {Yang}}, \bibinfo {author} {\bibfnamefont {Z.}~\bibnamefont {Su}}, \bibinfo {author} {\bibfnamefont {X.~Y.}\ \bibnamefont {Wang}}, \bibinfo {author} {\bibfnamefont {J.}~\bibnamefont {Rui}}, \bibinfo {author} {\bibfnamefont {B.}~\bibnamefont {Zhao}},\ and\ \bibinfo {author} {\bibfnamefont {J.~W.}\ \bibnamefont {Pan}},\ }\bibfield  {title} {\bibinfo {title} {{Preparation of a quantum degenerate mixture of Na 23 K 40 molecules and K 40 atoms}},\ }\href {https://doi.org/10.1103/PhysRevA.107.013307} {\bibfield  {journal} {\bibinfo  {journal} {Physical Review A}\ }\textbf {\bibinfo {volume} {107}},\ \bibinfo {pages} {1} (\bibinfo {year} {2023})}\BibitemShut {NoStop}%
\bibitem [{\citenamefont {Bigagli}\ \emph {et~al.}(2024)\citenamefont {Bigagli}, \citenamefont {Yuan}, \citenamefont {Zhang}, \citenamefont {Bulatovic}, \citenamefont {Karman}, \citenamefont {Stevenson},\ and\ \citenamefont {Will}}]{Bigagli2024}%
  \BibitemOpen
  \bibfield  {author} {\bibinfo {author} {\bibfnamefont {N.}~\bibnamefont {Bigagli}}, \bibinfo {author} {\bibfnamefont {W.}~\bibnamefont {Yuan}}, \bibinfo {author} {\bibfnamefont {S.}~\bibnamefont {Zhang}}, \bibinfo {author} {\bibfnamefont {B.}~\bibnamefont {Bulatovic}}, \bibinfo {author} {\bibfnamefont {T.}~\bibnamefont {Karman}}, \bibinfo {author} {\bibfnamefont {I.}~\bibnamefont {Stevenson}},\ and\ \bibinfo {author} {\bibfnamefont {S.}~\bibnamefont {Will}},\ }\bibfield  {title} {\bibinfo {title} {{Observation of Bose–Einstein condensation of dipolar molecules}},\ }\href {https://doi.org/10.1038/s41586-024-07492-z} {\bibfield  {journal} {\bibinfo  {journal} {Nature}\ }\textbf {\bibinfo {volume} {631}},\ \bibinfo {pages} {289} (\bibinfo {year} {2024})}\BibitemShut {NoStop}%
\bibitem [{\citenamefont {Danzl}\ \emph {et~al.}(2008)\citenamefont {Danzl}, \citenamefont {Haller}, \citenamefont {Gustavsson}, \citenamefont {Mark}, \citenamefont {Hart}, \citenamefont {Bouloufa}, \citenamefont {Dulieu}, \citenamefont {Ritsch},\ and\ \citenamefont {N{\"{a}}gerl}}]{Danzl2008}%
  \BibitemOpen
  \bibfield  {author} {\bibinfo {author} {\bibfnamefont {J.~G.}\ \bibnamefont {Danzl}}, \bibinfo {author} {\bibfnamefont {E.}~\bibnamefont {Haller}}, \bibinfo {author} {\bibfnamefont {M.}~\bibnamefont {Gustavsson}}, \bibinfo {author} {\bibfnamefont {M.~J.}\ \bibnamefont {Mark}}, \bibinfo {author} {\bibfnamefont {R.}~\bibnamefont {Hart}}, \bibinfo {author} {\bibfnamefont {N.}~\bibnamefont {Bouloufa}}, \bibinfo {author} {\bibfnamefont {O.}~\bibnamefont {Dulieu}}, \bibinfo {author} {\bibfnamefont {H.}~\bibnamefont {Ritsch}},\ and\ \bibinfo {author} {\bibfnamefont {H.-C.}\ \bibnamefont {N{\"{a}}gerl}},\ }\bibfield  {title} {\bibinfo {title} {{Quantum Gas of Deeply Bound Ground State Molecules}},\ }\href {https://doi.org/10.1126/science.1159909} {\bibfield  {journal} {\bibinfo  {journal} {Science}\ }\textbf {\bibinfo {volume} {321}},\ \bibinfo {pages} {1062} (\bibinfo {year} {2008})}\BibitemShut {NoStop}%
\bibitem [{\citenamefont {Stevenson}\ \emph {et~al.}(2023)\citenamefont {Stevenson}, \citenamefont {Lam}, \citenamefont {Bigagli}, \citenamefont {Warner}, \citenamefont {Yuan}, \citenamefont {Zhang},\ and\ \citenamefont {Will}}]{Stevenson2023}%
  \BibitemOpen
  \bibfield  {author} {\bibinfo {author} {\bibfnamefont {I.}~\bibnamefont {Stevenson}}, \bibinfo {author} {\bibfnamefont {A.~Z.}\ \bibnamefont {Lam}}, \bibinfo {author} {\bibfnamefont {N.}~\bibnamefont {Bigagli}}, \bibinfo {author} {\bibfnamefont {C.}~\bibnamefont {Warner}}, \bibinfo {author} {\bibfnamefont {W.}~\bibnamefont {Yuan}}, \bibinfo {author} {\bibfnamefont {S.}~\bibnamefont {Zhang}},\ and\ \bibinfo {author} {\bibfnamefont {S.}~\bibnamefont {Will}},\ }\bibfield  {title} {\bibinfo {title} {{Ultracold Gas of Dipolar NaCs Ground State Molecules}},\ }\href {https://doi.org/10.1103/PhysRevLett.130.113002} {\bibfield  {journal} {\bibinfo  {journal} {Physical Review Letters}\ }\textbf {\bibinfo {volume} {130}},\ \bibinfo {pages} {113002} (\bibinfo {year} {2023})}\BibitemShut {NoStop}%
\bibitem [{\citenamefont {Gregory}\ \emph {et~al.}(2019)\citenamefont {Gregory}, \citenamefont {Frye}, \citenamefont {Blackmore}, \citenamefont {Bridge}, \citenamefont {Sawant}, \citenamefont {Hutson},\ and\ \citenamefont {Cornish}}]{Gregory2019}%
  \BibitemOpen
  \bibfield  {author} {\bibinfo {author} {\bibfnamefont {P.~D.}\ \bibnamefont {Gregory}}, \bibinfo {author} {\bibfnamefont {M.~D.}\ \bibnamefont {Frye}}, \bibinfo {author} {\bibfnamefont {J.~A.}\ \bibnamefont {Blackmore}}, \bibinfo {author} {\bibfnamefont {E.~M.}\ \bibnamefont {Bridge}}, \bibinfo {author} {\bibfnamefont {R.}~\bibnamefont {Sawant}}, \bibinfo {author} {\bibfnamefont {J.~M.}\ \bibnamefont {Hutson}},\ and\ \bibinfo {author} {\bibfnamefont {S.~L.}\ \bibnamefont {Cornish}},\ }\bibfield  {title} {\bibinfo {title} {{Sticky collisions of ultracold RbCs molecules}},\ }\href {https://doi.org/10.1038/s41467-019-11033-y} {\bibfield  {journal} {\bibinfo  {journal} {Nature Communications}\ }\textbf {\bibinfo {volume} {10}},\ \bibinfo {pages} {3104} (\bibinfo {year} {2019})}\BibitemShut {NoStop}%
\bibitem [{\citenamefont {Gersema}\ \emph {et~al.}(2021)\citenamefont {Gersema}, \citenamefont {Voges}, \citenamefont {{Meyer zum Alten Borgloh}}, \citenamefont {Koch}, \citenamefont {Hartmann}, \citenamefont {Zenesini}, \citenamefont {Ospelkaus}, \citenamefont {Zenesini}, \citenamefont {Lin},\ and\ \citenamefont {He}}]{Gersema2021}%
  \BibitemOpen
  \bibfield  {author} {\bibinfo {author} {\bibfnamefont {P.}~\bibnamefont {Gersema}}, \bibinfo {author} {\bibfnamefont {K.~K.}\ \bibnamefont {Voges}}, \bibinfo {author} {\bibfnamefont {M.}~\bibnamefont {{Meyer zum Alten Borgloh}}}, \bibinfo {author} {\bibfnamefont {L.}~\bibnamefont {Koch}}, \bibinfo {author} {\bibfnamefont {T.}~\bibnamefont {Hartmann}}, \bibinfo {author} {\bibfnamefont {A.}~\bibnamefont {Zenesini}}, \bibinfo {author} {\bibfnamefont {S.}~\bibnamefont {Ospelkaus}}, \bibinfo {author} {\bibfnamefont {A.}~\bibnamefont {Zenesini}}, \bibinfo {author} {\bibfnamefont {J.}~\bibnamefont {Lin}},\ and\ \bibinfo {author} {\bibfnamefont {J.}~\bibnamefont {He}},\ }\bibfield  {title} {\bibinfo {title} {{Probing Photoinduced Two-Body Loss of Ultracold Nonreactive Bosonic and Molecules}},\ }\href {https://doi.org/10.1103/PhysRevLett.127.163401} {\bibfield  {journal} {\bibinfo  {journal} {Physical Review Letters}\ }\textbf {\bibinfo {volume} {127}},\ \bibinfo {pages} {163401} (\bibinfo {year}
  {2021})}\BibitemShut {NoStop}%
\bibitem [{\citenamefont {Bause}\ \emph {et~al.}(2021)\citenamefont {Bause}, \citenamefont {Schindewolf}, \citenamefont {Tao}, \citenamefont {Duda}, \citenamefont {Chen}, \citenamefont {Qu{\'{e}}m{\'{e}}ner}, \citenamefont {Karman}, \citenamefont {Christianen}, \citenamefont {Bloch},\ and\ \citenamefont {Luo}}]{Bause2021}%
  \BibitemOpen
  \bibfield  {author} {\bibinfo {author} {\bibfnamefont {R.}~\bibnamefont {Bause}}, \bibinfo {author} {\bibfnamefont {A.}~\bibnamefont {Schindewolf}}, \bibinfo {author} {\bibfnamefont {R.}~\bibnamefont {Tao}}, \bibinfo {author} {\bibfnamefont {M.}~\bibnamefont {Duda}}, \bibinfo {author} {\bibfnamefont {X.~Y.}\ \bibnamefont {Chen}}, \bibinfo {author} {\bibfnamefont {G.}~\bibnamefont {Qu{\'{e}}m{\'{e}}ner}}, \bibinfo {author} {\bibfnamefont {T.}~\bibnamefont {Karman}}, \bibinfo {author} {\bibfnamefont {A.}~\bibnamefont {Christianen}}, \bibinfo {author} {\bibfnamefont {I.}~\bibnamefont {Bloch}},\ and\ \bibinfo {author} {\bibfnamefont {X.~Y.}\ \bibnamefont {Luo}},\ }\bibfield  {title} {\bibinfo {title} {{Collisions of ultracold molecules in bright and dark optical dipole traps}},\ }\href {https://doi.org/10.1103/PhysRevResearch.3.033013} {\bibfield  {journal} {\bibinfo  {journal} {Physical Review Research}\ }\textbf {\bibinfo {volume} {3}},\ \bibinfo {pages} {033013} (\bibinfo {year} {2021})}\BibitemShut
  {NoStop}%
\bibitem [{\citenamefont {Gregory}\ \emph {et~al.}(2020)\citenamefont {Gregory}, \citenamefont {Blackmore}, \citenamefont {Bromley},\ and\ \citenamefont {Cornish}}]{Gregory2020}%
  \BibitemOpen
  \bibfield  {author} {\bibinfo {author} {\bibfnamefont {P.~D.}\ \bibnamefont {Gregory}}, \bibinfo {author} {\bibfnamefont {J.~A.}\ \bibnamefont {Blackmore}}, \bibinfo {author} {\bibfnamefont {S.~L.}\ \bibnamefont {Bromley}},\ and\ \bibinfo {author} {\bibfnamefont {S.~L.}\ \bibnamefont {Cornish}},\ }\bibfield  {title} {\bibinfo {title} {{Loss of Ultracold $^{87}$Rb$^{133}$Cs Molecules via Optical Excitation of Long-Lived Two-Body Collision Complexes}},\ }\href {https://doi.org/10.1103/PhysRevLett.124.163402} {\bibfield  {journal} {\bibinfo  {journal} {Physical Review Letters}\ }\textbf {\bibinfo {volume} {124}},\ \bibinfo {pages} {163402} (\bibinfo {year} {2020})}\BibitemShut {NoStop}%
\bibitem [{\citenamefont {Liu}\ \emph {et~al.}(2020)\citenamefont {Liu}, \citenamefont {Hu}, \citenamefont {Nichols}, \citenamefont {Grimes}, \citenamefont {Karman}, \citenamefont {Guo},\ and\ \citenamefont {Ni}}]{Liu2020}%
  \BibitemOpen
  \bibfield  {author} {\bibinfo {author} {\bibfnamefont {Y.}~\bibnamefont {Liu}}, \bibinfo {author} {\bibfnamefont {M.~G.}\ \bibnamefont {Hu}}, \bibinfo {author} {\bibfnamefont {M.~A.}\ \bibnamefont {Nichols}}, \bibinfo {author} {\bibfnamefont {D.~D.}\ \bibnamefont {Grimes}}, \bibinfo {author} {\bibfnamefont {T.}~\bibnamefont {Karman}}, \bibinfo {author} {\bibfnamefont {H.}~\bibnamefont {Guo}},\ and\ \bibinfo {author} {\bibfnamefont {K.~K.}\ \bibnamefont {Ni}},\ }\bibfield  {title} {\bibinfo {title} {{Photo-excitation of long-lived transient intermediates in ultracold reactions}},\ }\href {https://doi.org/10.1038/s41567-020-0968-8} {\bibfield  {journal} {\bibinfo  {journal} {Nature Physics}\ }\textbf {\bibinfo {volume} {16}},\ \bibinfo {pages} {1132} (\bibinfo {year} {2020})}\BibitemShut {NoStop}%
\bibitem [{\citenamefont {Bause}\ \emph {et~al.}(2023)\citenamefont {Bause}, \citenamefont {Christianen}, \citenamefont {Schindewolf}, \citenamefont {Bloch},\ and\ \citenamefont {Luo}}]{Bause2023}%
  \BibitemOpen
  \bibfield  {author} {\bibinfo {author} {\bibfnamefont {R.}~\bibnamefont {Bause}}, \bibinfo {author} {\bibfnamefont {A.}~\bibnamefont {Christianen}}, \bibinfo {author} {\bibfnamefont {A.}~\bibnamefont {Schindewolf}}, \bibinfo {author} {\bibfnamefont {I.}~\bibnamefont {Bloch}},\ and\ \bibinfo {author} {\bibfnamefont {X.~Y.}\ \bibnamefont {Luo}},\ }\bibfield  {title} {\bibinfo {title} {{Ultracold Sticky Collisions: Theoretical and Experimental Status}},\ }\href {https://doi.org/10.1021/acs.jpca.2c08095} {\bibfield  {journal} {\bibinfo  {journal} {Journal of Physical Chemistry A}\ }\textbf {\bibinfo {volume} {127}},\ \bibinfo {pages} {729} (\bibinfo {year} {2023})}\BibitemShut {NoStop}%
\bibitem [{\citenamefont {Valtolina}\ \emph {et~al.}(2020)\citenamefont {Valtolina}, \citenamefont {Matsuda}, \citenamefont {Tobias}, \citenamefont {Li}, \citenamefont {de~Marco},\ and\ \citenamefont {Ye}}]{Valtolina.2020}%
  \BibitemOpen
  \bibfield  {author} {\bibinfo {author} {\bibfnamefont {G.}~\bibnamefont {Valtolina}}, \bibinfo {author} {\bibfnamefont {K.}~\bibnamefont {Matsuda}}, \bibinfo {author} {\bibfnamefont {W.~G.}\ \bibnamefont {Tobias}}, \bibinfo {author} {\bibfnamefont {J.-R.}\ \bibnamefont {Li}}, \bibinfo {author} {\bibfnamefont {L.}~\bibnamefont {de~Marco}},\ and\ \bibinfo {author} {\bibfnamefont {J.}~\bibnamefont {Ye}},\ }\bibfield  {title} {\bibinfo {title} {{Dipolar evaporation of reactive molecules to below the Fermi temperature}},\ }\href {https://doi.org/10.1038/s41586-020-2980-7} {\bibfield  {journal} {\bibinfo  {journal} {Nature}\ }\textbf {\bibinfo {volume} {588}},\ \bibinfo {pages} {239} (\bibinfo {year} {2020})}\BibitemShut {NoStop}%
\bibitem [{\citenamefont {Li}\ \emph {et~al.}(2021)\citenamefont {Li}, \citenamefont {Tobias}, \citenamefont {Matsuda}, \citenamefont {Miller}, \citenamefont {Valtolina}, \citenamefont {{De Marco}}, \citenamefont {Wang}, \citenamefont {Lassabli{\`{e}}re}, \citenamefont {Qu{\'{e}}m{\'{e}}ner}, \citenamefont {Bohn},\ and\ \citenamefont {Ye}}]{Li2021}%
  \BibitemOpen
  \bibfield  {author} {\bibinfo {author} {\bibfnamefont {J.~R.}\ \bibnamefont {Li}}, \bibinfo {author} {\bibfnamefont {W.~G.}\ \bibnamefont {Tobias}}, \bibinfo {author} {\bibfnamefont {K.}~\bibnamefont {Matsuda}}, \bibinfo {author} {\bibfnamefont {C.}~\bibnamefont {Miller}}, \bibinfo {author} {\bibfnamefont {G.}~\bibnamefont {Valtolina}}, \bibinfo {author} {\bibfnamefont {L.}~\bibnamefont {{De Marco}}}, \bibinfo {author} {\bibfnamefont {R.~R.}\ \bibnamefont {Wang}}, \bibinfo {author} {\bibfnamefont {L.}~\bibnamefont {Lassabli{\`{e}}re}}, \bibinfo {author} {\bibfnamefont {G.}~\bibnamefont {Qu{\'{e}}m{\'{e}}ner}}, \bibinfo {author} {\bibfnamefont {J.~L.}\ \bibnamefont {Bohn}},\ and\ \bibinfo {author} {\bibfnamefont {J.}~\bibnamefont {Ye}},\ }\bibfield  {title} {\bibinfo {title} {{Tuning of dipolar interactions and evaporative cooling in a three-dimensional molecular quantum gas}},\ }\href {https://doi.org/10.1038/s41567-021-01329-6} {\bibfield  {journal} {\bibinfo  {journal} {Nature Physics}\ }\textbf {\bibinfo
  {volume} {17}},\ \bibinfo {pages} {1144} (\bibinfo {year} {2021})}\BibitemShut {NoStop}%
\bibitem [{\citenamefont {Anderegg}\ \emph {et~al.}(2021)\citenamefont {Anderegg}, \citenamefont {Burchesky}, \citenamefont {Bao}, \citenamefont {Yu}, \citenamefont {Karman}, \citenamefont {Chae}, \citenamefont {Ni}, \citenamefont {Ketterle},\ and\ \citenamefont {Doyle}}]{Anderegg2021}%
  \BibitemOpen
  \bibfield  {author} {\bibinfo {author} {\bibfnamefont {L.}~\bibnamefont {Anderegg}}, \bibinfo {author} {\bibfnamefont {S.}~\bibnamefont {Burchesky}}, \bibinfo {author} {\bibfnamefont {Y.}~\bibnamefont {Bao}}, \bibinfo {author} {\bibfnamefont {S.~S.}\ \bibnamefont {Yu}}, \bibinfo {author} {\bibfnamefont {T.}~\bibnamefont {Karman}}, \bibinfo {author} {\bibfnamefont {E.}~\bibnamefont {Chae}}, \bibinfo {author} {\bibfnamefont {K.~K.}\ \bibnamefont {Ni}}, \bibinfo {author} {\bibfnamefont {W.}~\bibnamefont {Ketterle}},\ and\ \bibinfo {author} {\bibfnamefont {J.~M.}\ \bibnamefont {Doyle}},\ }\bibfield  {title} {\bibinfo {title} {{Observation of microwave shielding of ultracold molecules}},\ }\href {https://doi.org/10.1126/science.abg9502} {\bibfield  {journal} {\bibinfo  {journal} {Science}\ }\textbf {\bibinfo {volume} {373}},\ \bibinfo {pages} {779} (\bibinfo {year} {2021})}\BibitemShut {NoStop}%
\bibitem [{\citenamefont {Schindewolf}\ \emph {et~al.}(2022)\citenamefont {Schindewolf}, \citenamefont {Bause}, \citenamefont {Chen}, \citenamefont {Duda}, \citenamefont {Karman}, \citenamefont {Bloch},\ and\ \citenamefont {Luo}}]{Schindewolf2022}%
  \BibitemOpen
  \bibfield  {author} {\bibinfo {author} {\bibfnamefont {A.}~\bibnamefont {Schindewolf}}, \bibinfo {author} {\bibfnamefont {R.}~\bibnamefont {Bause}}, \bibinfo {author} {\bibfnamefont {X.~Y.}\ \bibnamefont {Chen}}, \bibinfo {author} {\bibfnamefont {M.}~\bibnamefont {Duda}}, \bibinfo {author} {\bibfnamefont {T.}~\bibnamefont {Karman}}, \bibinfo {author} {\bibfnamefont {I.}~\bibnamefont {Bloch}},\ and\ \bibinfo {author} {\bibfnamefont {X.~Y.}\ \bibnamefont {Luo}},\ }\bibfield  {title} {\bibinfo {title} {{Evaporation of microwave-shielded polar molecules to quantum degeneracy}},\ }\href {https://doi.org/10.1038/s41586-022-04900-0} {\bibfield  {journal} {\bibinfo  {journal} {Nature}\ }\textbf {\bibinfo {volume} {607}},\ \bibinfo {pages} {677} (\bibinfo {year} {2022})}\BibitemShut {NoStop}%
\bibitem [{\citenamefont {Lin}\ \emph {et~al.}(2023)\citenamefont {Lin}, \citenamefont {Chen}, \citenamefont {Jin}, \citenamefont {Shi}, \citenamefont {Deng}, \citenamefont {Zhang}, \citenamefont {Qu{\'{e}}m{\'{e}}ner}, \citenamefont {Shi}, \citenamefont {Yi},\ and\ \citenamefont {Wang}}]{Lin2023}%
  \BibitemOpen
  \bibfield  {author} {\bibinfo {author} {\bibfnamefont {J.}~\bibnamefont {Lin}}, \bibinfo {author} {\bibfnamefont {G.}~\bibnamefont {Chen}}, \bibinfo {author} {\bibfnamefont {M.}~\bibnamefont {Jin}}, \bibinfo {author} {\bibfnamefont {Z.}~\bibnamefont {Shi}}, \bibinfo {author} {\bibfnamefont {F.}~\bibnamefont {Deng}}, \bibinfo {author} {\bibfnamefont {W.}~\bibnamefont {Zhang}}, \bibinfo {author} {\bibfnamefont {G.}~\bibnamefont {Qu{\'{e}}m{\'{e}}ner}}, \bibinfo {author} {\bibfnamefont {T.}~\bibnamefont {Shi}}, \bibinfo {author} {\bibfnamefont {S.}~\bibnamefont {Yi}},\ and\ \bibinfo {author} {\bibfnamefont {D.}~\bibnamefont {Wang}},\ }\bibfield  {title} {\bibinfo {title} {{Microwave Shielding of Bosonic NaRb Molecules}},\ }\href {https://doi.org/10.1103/physrevx.13.031032} {\bibfield  {journal} {\bibinfo  {journal} {Physical Review X}\ }\textbf {\bibinfo {volume} {13}},\ \bibinfo {pages} {031032} (\bibinfo {year} {2023})}\BibitemShut {NoStop}%
\bibitem [{\citenamefont {Bigagli}\ \emph {et~al.}(2023)\citenamefont {Bigagli}, \citenamefont {Warner}, \citenamefont {Yuan}, \citenamefont {Zhang}, \citenamefont {Stevenson}, \citenamefont {Karman},\ and\ \citenamefont {Will}}]{Bigagli2023}%
  \BibitemOpen
  \bibfield  {author} {\bibinfo {author} {\bibfnamefont {N.}~\bibnamefont {Bigagli}}, \bibinfo {author} {\bibfnamefont {C.}~\bibnamefont {Warner}}, \bibinfo {author} {\bibfnamefont {W.}~\bibnamefont {Yuan}}, \bibinfo {author} {\bibfnamefont {S.}~\bibnamefont {Zhang}}, \bibinfo {author} {\bibfnamefont {I.}~\bibnamefont {Stevenson}}, \bibinfo {author} {\bibfnamefont {T.}~\bibnamefont {Karman}},\ and\ \bibinfo {author} {\bibfnamefont {S.}~\bibnamefont {Will}},\ }\bibfield  {title} {\bibinfo {title} {{Collisionally stable gas of bosonic dipolar ground-state molecules}},\ }\href {https://doi.org/10.1038/s41567-023-02200-6} {\bibfield  {journal} {\bibinfo  {journal} {Nature Physics}\ }\textbf {\bibinfo {volume} {19}},\ \bibinfo {pages} {1579} (\bibinfo {year} {2023})}\BibitemShut {NoStop}%
\bibitem [{\citenamefont {Ladjimi}\ and\ \citenamefont {Tomza}(2024)}]{Ladjimi2024}%
  \BibitemOpen
  \bibfield  {author} {\bibinfo {author} {\bibfnamefont {H.}~\bibnamefont {Ladjimi}}\ and\ \bibinfo {author} {\bibfnamefont {M.}~\bibnamefont {Tomza}},\ }\bibfield  {title} {\bibinfo {title} {{Diatomic molecules of alkali-metal and alkaline-earth-metal atoms: Interaction potentials, dipole moments, and polarizabilities}},\ }\href {https://doi.org/10.1103/PhysRevA.109.052814} {\bibfield  {journal} {\bibinfo  {journal} {Physical Review A}\ }\textbf {\bibinfo {volume} {109}},\ \bibinfo {pages} {52814} (\bibinfo {year} {2024})}\BibitemShut {NoStop}%
\bibitem [{\citenamefont {Vexiau}\ \emph {et~al.}(2017)\citenamefont {Vexiau}, \citenamefont {Borsalino}, \citenamefont {Lepers}, \citenamefont {Orb{\'{a}}n}, \citenamefont {Aymar}, \citenamefont {Dulieu},\ and\ \citenamefont {Bouloufa-Maafa}}]{Vexiau2017}%
  \BibitemOpen
  \bibfield  {author} {\bibinfo {author} {\bibfnamefont {R.}~\bibnamefont {Vexiau}}, \bibinfo {author} {\bibfnamefont {D.}~\bibnamefont {Borsalino}}, \bibinfo {author} {\bibfnamefont {M.}~\bibnamefont {Lepers}}, \bibinfo {author} {\bibfnamefont {A.}~\bibnamefont {Orb{\'{a}}n}}, \bibinfo {author} {\bibfnamefont {M.}~\bibnamefont {Aymar}}, \bibinfo {author} {\bibfnamefont {O.}~\bibnamefont {Dulieu}},\ and\ \bibinfo {author} {\bibfnamefont {N.}~\bibnamefont {Bouloufa-Maafa}},\ }\bibfield  {title} {\bibinfo {title} {{Dynamic dipole polarizabilities of heteronuclear alkali dimers: Optical response, trapping and control of ultracold molecules}},\ }\href {https://doi.org/10.1080/0144235X.2017.1351821} {\bibfield  {journal} {\bibinfo  {journal} {International Reviews in Physical Chemistry}\ }\textbf {\bibinfo {volume} {36}},\ \bibinfo {pages} {709} (\bibinfo {year} {2017})}\BibitemShut {NoStop}%
\bibitem [{Cha()}]{Charly}%
  \BibitemOpen
  \bibinfo {title} {{Manuscript in preparation}}\BibitemShut {NoStop}%
\bibitem [{sup()}]{supp}%
  \BibitemOpen
\bibfield  {title} {  }\bibinfo {title} {{Supplemental Material}}\BibitemShut {NoStop}%
\bibitem [{\citenamefont {Borsalino}\ \emph {et~al.}(2016)\citenamefont {Borsalino}, \citenamefont {Vexiau}, \citenamefont {Aymar}, \citenamefont {Luc-Koenig}, \citenamefont {Dulieu},\ and\ \citenamefont {Bouloufa-Maafa}}]{Borsalino2016}%
  \BibitemOpen
\bibfield  {title} {  }\bibfield  {author} {\bibinfo {author} {\bibfnamefont {D.}~\bibnamefont {Borsalino}}, \bibinfo {author} {\bibfnamefont {R.}~\bibnamefont {Vexiau}}, \bibinfo {author} {\bibfnamefont {M.}~\bibnamefont {Aymar}}, \bibinfo {author} {\bibfnamefont {E.}~\bibnamefont {Luc-Koenig}}, \bibinfo {author} {\bibfnamefont {O.}~\bibnamefont {Dulieu}},\ and\ \bibinfo {author} {\bibfnamefont {N.}~\bibnamefont {Bouloufa-Maafa}},\ }\bibfield  {title} {\bibinfo {title} {{Prospects for the formation of ultracold polar ground state KCs molecules via an optical process}},\ }\href {https://doi.org/10.1088/0953-4075/49/5/055301} {\bibfield  {journal} {\bibinfo  {journal} {Journal of Physics B: Atomic, Molecular and Optical Physics}\ }\textbf {\bibinfo {volume} {49}},\ \bibinfo {pages} {055301} (\bibinfo {year} {2016})}\BibitemShut {NoStop}%
\bibitem [{\citenamefont {Drever}\ \emph {et~al.}(1983)\citenamefont {Drever}, \citenamefont {Hall}, \citenamefont {Kowalski}, \citenamefont {Hough}, \citenamefont {Ford}, \citenamefont {Munley},\ and\ \citenamefont {Ward}}]{Drever1983}%
  \BibitemOpen
  \bibfield  {author} {\bibinfo {author} {\bibfnamefont {R.~W.~P.}\ \bibnamefont {Drever}}, \bibinfo {author} {\bibfnamefont {J.~L.}\ \bibnamefont {Hall}}, \bibinfo {author} {\bibfnamefont {F.~V.}\ \bibnamefont {Kowalski}}, \bibinfo {author} {\bibfnamefont {J.}~\bibnamefont {Hough}}, \bibinfo {author} {\bibfnamefont {G.~M.}\ \bibnamefont {Ford}}, \bibinfo {author} {\bibfnamefont {A.~J.}\ \bibnamefont {Munley}},\ and\ \bibinfo {author} {\bibfnamefont {H.}~\bibnamefont {Ward}},\ }\bibfield  {title} {\bibinfo {title} {{Laser phase and frequency stabilization using an optical resonator}},\ }\href {https://doi.org/10.1007/BF00702605} {\bibfield  {journal} {\bibinfo  {journal} {Applied Physics B Photophysics and Laser Chemistry}\ }\textbf {\bibinfo {volume} {31}},\ \bibinfo {pages} {97} (\bibinfo {year} {1983})}\BibitemShut {NoStop}%
\bibitem [{Nad()}]{Nadia}%
  \BibitemOpen
  \bibinfo {title} {{Private communication with Nadia Bouloufa-Maafa}}\BibitemShut {NoStop}%
\bibitem [{\citenamefont {Vitanov}\ and\ \citenamefont {Stenholm}(1999)}]{Vitanov1999}%
  \BibitemOpen
\bibfield  {title} {  }\bibfield  {author} {\bibinfo {author} {\bibfnamefont {N.~V.}\ \bibnamefont {Vitanov}}\ and\ \bibinfo {author} {\bibfnamefont {S.}~\bibnamefont {Stenholm}},\ }\bibfield  {title} {\bibinfo {title} {{Adiabatic population transfer via multiple intermediate states}},\ }\href {https://doi.org/10.1103/PhysRevA.60.3820} {\bibfield  {journal} {\bibinfo  {journal} {Physical Review A}\ }\textbf {\bibinfo {volume} {60}},\ \bibinfo {pages} {3820} (\bibinfo {year} {1999})}\BibitemShut {NoStop}%
\bibitem [{\citenamefont {Guo}\ \emph {et~al.}(2017)\citenamefont {Guo}, \citenamefont {Vexiau}, \citenamefont {Zhu}, \citenamefont {Lu}, \citenamefont {Bouloufa-Maafa}, \citenamefont {Dulieu},\ and\ \citenamefont {Wang}}]{Guo2017}%
  \BibitemOpen
  \bibfield  {author} {\bibinfo {author} {\bibfnamefont {M.}~\bibnamefont {Guo}}, \bibinfo {author} {\bibfnamefont {R.}~\bibnamefont {Vexiau}}, \bibinfo {author} {\bibfnamefont {B.}~\bibnamefont {Zhu}}, \bibinfo {author} {\bibfnamefont {B.}~\bibnamefont {Lu}}, \bibinfo {author} {\bibfnamefont {N.}~\bibnamefont {Bouloufa-Maafa}}, \bibinfo {author} {\bibfnamefont {O.}~\bibnamefont {Dulieu}},\ and\ \bibinfo {author} {\bibfnamefont {D.}~\bibnamefont {Wang}},\ }\bibfield  {title} {\bibinfo {title} {{High-resolution molecular spectroscopy for producing ultracold absolute-ground-state $^{23}$Na$^{87}$Rb molecules}},\ }\href {https://doi.org/10.1103/PhysRevA.96.052505} {\bibfield  {journal} {\bibinfo  {journal} {Physical Review A}\ }\textbf {\bibinfo {volume} {96}},\ \bibinfo {pages} {052505} (\bibinfo {year} {2017})}\BibitemShut {NoStop}%
\bibitem [{\citenamefont {Tamanis}\ \emph {et~al.}(2010)\citenamefont {Tamanis}, \citenamefont {Klincare}, \citenamefont {Kruzins}, \citenamefont {Nikolayeva}, \citenamefont {Ferber}, \citenamefont {Pazyuk},\ and\ \citenamefont {Stolyarov}}]{Tamanis2010}%
  \BibitemOpen
  \bibfield  {author} {\bibinfo {author} {\bibfnamefont {M.}~\bibnamefont {Tamanis}}, \bibinfo {author} {\bibfnamefont {I.}~\bibnamefont {Klincare}}, \bibinfo {author} {\bibfnamefont {A.}~\bibnamefont {Kruzins}}, \bibinfo {author} {\bibfnamefont {O.}~\bibnamefont {Nikolayeva}}, \bibinfo {author} {\bibfnamefont {R.}~\bibnamefont {Ferber}}, \bibinfo {author} {\bibfnamefont {E.~A.}\ \bibnamefont {Pazyuk}},\ and\ \bibinfo {author} {\bibfnamefont {A.~V.}\ \bibnamefont {Stolyarov}},\ }\bibfield  {title} {\bibinfo {title} {{Direct excitation of the "dark" $b^3\Pi$ state predicted by deperturbation analysis of the $A^1\Sigma^+-b^3\Pi$ complex in KCs}},\ }\href {https://doi.org/10.1103/PhysRevA.82.032506} {\bibfield  {journal} {\bibinfo  {journal} {Physical Review A}\ }\textbf {\bibinfo {volume} {82}},\ \bibinfo {pages} {032506} (\bibinfo {year} {2010})}\BibitemShut {NoStop}%
\bibitem [{\citenamefont {Kruzins}\ \emph {et~al.}(2010)\citenamefont {Kruzins}, \citenamefont {Klincare}, \citenamefont {Nikolayeva}, \citenamefont {Tamanis}, \citenamefont {Ferber}, \citenamefont {Pazyuk},\ and\ \citenamefont {Stolyarov}}]{Kruzins2010}%
  \BibitemOpen
  \bibfield  {author} {\bibinfo {author} {\bibfnamefont {A.}~\bibnamefont {Kruzins}}, \bibinfo {author} {\bibfnamefont {I.}~\bibnamefont {Klincare}}, \bibinfo {author} {\bibfnamefont {O.}~\bibnamefont {Nikolayeva}}, \bibinfo {author} {\bibfnamefont {M.}~\bibnamefont {Tamanis}}, \bibinfo {author} {\bibfnamefont {R.}~\bibnamefont {Ferber}}, \bibinfo {author} {\bibfnamefont {E.~A.}\ \bibnamefont {Pazyuk}},\ and\ \bibinfo {author} {\bibfnamefont {A.~V.}\ \bibnamefont {Stolyarov}},\ }\bibfield  {title} {\bibinfo {title} {{Fourier-transform spectroscopy and coupled-channels deperturbation treatment of the $A^1\Sigma^+-b^3\Pi$ complex of KCs}},\ }\href {https://doi.org/10.1103/PhysRevA.81.042509} {\bibfield  {journal} {\bibinfo  {journal} {Physical Review A}\ }\textbf {\bibinfo {volume} {81}},\ \bibinfo {pages} {042509} (\bibinfo {year} {2010})}\BibitemShut {NoStop}%
\bibitem [{\citenamefont {Kruzins}\ \emph {et~al.}(2013)\citenamefont {Kruzins}, \citenamefont {Klincare}, \citenamefont {Nikolayeva}, \citenamefont {Tamanis}, \citenamefont {Ferber}, \citenamefont {Pazyuk},\ and\ \citenamefont {Stolyarov}}]{Kruzins2013}%
  \BibitemOpen
  \bibfield  {author} {\bibinfo {author} {\bibfnamefont {A.}~\bibnamefont {Kruzins}}, \bibinfo {author} {\bibfnamefont {I.}~\bibnamefont {Klincare}}, \bibinfo {author} {\bibfnamefont {O.}~\bibnamefont {Nikolayeva}}, \bibinfo {author} {\bibfnamefont {M.}~\bibnamefont {Tamanis}}, \bibinfo {author} {\bibfnamefont {R.}~\bibnamefont {Ferber}}, \bibinfo {author} {\bibfnamefont {E.~A.}\ \bibnamefont {Pazyuk}},\ and\ \bibinfo {author} {\bibfnamefont {A.~V.}\ \bibnamefont {Stolyarov}},\ }\bibfield  {title} {\bibinfo {title} {{Fourier-transform spectroscopy of $(4)^1\Sigma^+$ → $A^1\Sigma^+-b^3\Pi$, $A^1\Sigma^+ - b^3\Pi$ → $X^1\Sigma^+$, and $(1)^3\Delta_1$→$b\Pi_{0±}$ transitions in KCs and deperturbation treatment of $A^1\Sigma^+$ and $b^3\Pi$ states}},\ }\href {https://doi.org/10.1063/1.4844275} {\bibfield  {journal} {\bibinfo  {journal} {Journal of Chemical Physics}\ }\textbf {\bibinfo {volume} {139}},\ \bibinfo {pages} {244301} (\bibinfo {year} {2013})}\BibitemShut {NoStop}%
\bibitem [{\citenamefont {Debatin}\ \emph {et~al.}(2011)\citenamefont {Debatin}, \citenamefont {Takekoshi}, \citenamefont {Rameshan}, \citenamefont {Reichs{\"{o}}llner}, \citenamefont {Ferlaino}, \citenamefont {Grimm}, \citenamefont {Vexiau}, \citenamefont {Bouloufa}, \citenamefont {Dulieu},\ and\ \citenamefont {N{\"{a}}gerl}}]{Debatin2011}%
  \BibitemOpen
  \bibfield  {author} {\bibinfo {author} {\bibfnamefont {M.}~\bibnamefont {Debatin}}, \bibinfo {author} {\bibfnamefont {T.}~\bibnamefont {Takekoshi}}, \bibinfo {author} {\bibfnamefont {R.}~\bibnamefont {Rameshan}}, \bibinfo {author} {\bibfnamefont {L.}~\bibnamefont {Reichs{\"{o}}llner}}, \bibinfo {author} {\bibfnamefont {F.}~\bibnamefont {Ferlaino}}, \bibinfo {author} {\bibfnamefont {R.}~\bibnamefont {Grimm}}, \bibinfo {author} {\bibfnamefont {R.}~\bibnamefont {Vexiau}}, \bibinfo {author} {\bibfnamefont {N.}~\bibnamefont {Bouloufa}}, \bibinfo {author} {\bibfnamefont {O.}~\bibnamefont {Dulieu}},\ and\ \bibinfo {author} {\bibfnamefont {H.-C.}\ \bibnamefont {N{\"{a}}gerl}},\ }\bibfield  {title} {\bibinfo {title} {{Molecular spectroscopy for ground-state transfer of ultracold RbCs molecules}},\ }\href {https://doi.org/10.1039/c1cp21769k} {\bibfield  {journal} {\bibinfo  {journal} {Physical Chemistry Chemical Physics}\ }\textbf {\bibinfo {volume} {13}},\ \bibinfo {pages} {18926} (\bibinfo {year}
  {2011})}\BibitemShut {NoStop}%
\bibitem [{\citenamefont {Ferber}\ \emph {et~al.}(2013)\citenamefont {Ferber}, \citenamefont {Nikolayeva}, \citenamefont {Tamanis}, \citenamefont {Kn{\"{o}}ckel},\ and\ \citenamefont {Tiemann}}]{Ferber2013}%
  \BibitemOpen
  \bibfield  {author} {\bibinfo {author} {\bibfnamefont {R.}~\bibnamefont {Ferber}}, \bibinfo {author} {\bibfnamefont {O.}~\bibnamefont {Nikolayeva}}, \bibinfo {author} {\bibfnamefont {M.}~\bibnamefont {Tamanis}}, \bibinfo {author} {\bibfnamefont {H.}~\bibnamefont {Kn{\"{o}}ckel}},\ and\ \bibinfo {author} {\bibfnamefont {E.}~\bibnamefont {Tiemann}},\ }\bibfield  {title} {\bibinfo {title} {{Long-range coupling of $X^1\Sigma^+$ and $a^3\Sigma^+$ states of the atom pair K+Cs}},\ }\href {https://doi.org/10.1103/PhysRevA.88.012516} {\bibfield  {journal} {\bibinfo  {journal} {Physical Review A}\ }\textbf {\bibinfo {volume} {88}},\ \bibinfo {pages} {012516} (\bibinfo {year} {2013})}\BibitemShut {NoStop}%
\bibitem [{\citenamefont {Fleischhauer}\ \emph {et~al.}(2005)\citenamefont {Fleischhauer}, \citenamefont {Imamoglu},\ and\ \citenamefont {Marangos}}]{Fleischhauer2005}%
  \BibitemOpen
  \bibfield  {author} {\bibinfo {author} {\bibfnamefont {M.}~\bibnamefont {Fleischhauer}}, \bibinfo {author} {\bibfnamefont {A.}~\bibnamefont {Imamoglu}},\ and\ \bibinfo {author} {\bibfnamefont {P.~J.}\ \bibnamefont {Marangos}},\ }\bibfield  {title} {\bibinfo {title} {{Electromagnetically induced transparency}},\ }\href {https://doi.org/10.1103/RevModPhys.77.633} {\bibfield  {journal} {\bibinfo  {journal} {Reviews of Modern Physics}\ }\textbf {\bibinfo {volume} {77}},\ \bibinfo {pages} {633} (\bibinfo {year} {2005})}\BibitemShut {NoStop}%
\bibitem [{\citenamefont {Yatsenko}\ \emph {et~al.}(2014)\citenamefont {Yatsenko}, \citenamefont {Shore},\ and\ \citenamefont {Bergmann}}]{Yatsenko2014}%
  \BibitemOpen
  \bibfield  {author} {\bibinfo {author} {\bibfnamefont {L.~P.}\ \bibnamefont {Yatsenko}}, \bibinfo {author} {\bibfnamefont {B.~W.}\ \bibnamefont {Shore}},\ and\ \bibinfo {author} {\bibfnamefont {K.}~\bibnamefont {Bergmann}},\ }\bibfield  {title} {\bibinfo {title} {{Detrimental consequences of small rapid laser fluctuations on stimulated Raman adiabatic passage}},\ }\href {https://doi.org/10.1103/PhysRevA.89.013831} {\bibfield  {journal} {\bibinfo  {journal} {Physical Review A}\ }\textbf {\bibinfo {volume} {89}},\ \bibinfo {pages} {013831} (\bibinfo {year} {2014})}\BibitemShut {NoStop}%
\bibitem [{\citenamefont {Julienne}\ \emph {et~al.}(2011)\citenamefont {Julienne}, \citenamefont {Hanna},\ and\ \citenamefont {Idziaszek}}]{Julienne2011}%
  \BibitemOpen
  \bibfield  {author} {\bibinfo {author} {\bibfnamefont {P.~S.}\ \bibnamefont {Julienne}}, \bibinfo {author} {\bibfnamefont {T.~M.}\ \bibnamefont {Hanna}},\ and\ \bibinfo {author} {\bibfnamefont {Z.}~\bibnamefont {Idziaszek}},\ }\bibfield  {title} {\bibinfo {title} {{Universal ultracold collision rates for polar molecules of two alkali-metal atoms}},\ }\href {https://doi.org/10.1039/c1cp21270b} {\bibfield  {journal} {\bibinfo  {journal} {Physical Chemistry Chemical Physics}\ }\textbf {\bibinfo {volume} {13}},\ \bibinfo {pages} {19114} (\bibinfo {year} {2011})}\BibitemShut {NoStop}%
\bibitem [{\citenamefont {Aldegunde}\ and\ \citenamefont {Hutson}(2017)}]{Aldegunde2017}%
  \BibitemOpen
  \bibfield  {author} {\bibinfo {author} {\bibfnamefont {J.}~\bibnamefont {Aldegunde}}\ and\ \bibinfo {author} {\bibfnamefont {J.~M.}\ \bibnamefont {Hutson}},\ }\bibfield  {title} {\bibinfo {title} {{Hyperfine structure of alkali-metal diatomic molecules}},\ }\href {https://doi.org/10.1103/PhysRevA.96.042506} {\bibfield  {journal} {\bibinfo  {journal} {Physical Review A}\ }\textbf {\bibinfo {volume} {96}},\ \bibinfo {pages} {042506} (\bibinfo {year} {2017})}\BibitemShut {NoStop}%
\bibitem [{\citenamefont {Aldegunde}\ \emph {et~al.}(2008)\citenamefont {Aldegunde}, \citenamefont {Rivington}, \citenamefont {Zuchowski},\ and\ \citenamefont {Hutson}}]{Aldegunde2008}%
  \BibitemOpen
  \bibfield  {author} {\bibinfo {author} {\bibfnamefont {J.}~\bibnamefont {Aldegunde}}, \bibinfo {author} {\bibfnamefont {B.~A.}\ \bibnamefont {Rivington}}, \bibinfo {author} {\bibfnamefont {P.~S.}\ \bibnamefont {Zuchowski}},\ and\ \bibinfo {author} {\bibfnamefont {J.~M.}\ \bibnamefont {Hutson}},\ }\bibfield  {title} {\bibinfo {title} {{Hyperfine energy levels of alkali-metal dimers: Ground-state polar molecules in electric and magnetic fields}},\ }\href {https://doi.org/10.1103/PhysRevA.78.033434} {\bibfield  {journal} {\bibinfo  {journal} {Physical Review A}\ }\textbf {\bibinfo {volume} {78}},\ \bibinfo {pages} {033434} (\bibinfo {year} {2008})}\BibitemShut {NoStop}%
\bibitem [{\citenamefont {Reinaudi}\ \emph {et~al.}(2007)\citenamefont {Reinaudi}, \citenamefont {Lahaye}, \citenamefont {Wang},\ and\ \citenamefont {Gu\'{e}ry-Odelin}}]{Reinaudi:07}%
  \BibitemOpen
  \bibfield  {author} {\bibinfo {author} {\bibfnamefont {G.}~\bibnamefont {Reinaudi}}, \bibinfo {author} {\bibfnamefont {T.}~\bibnamefont {Lahaye}}, \bibinfo {author} {\bibfnamefont {Z.}~\bibnamefont {Wang}},\ and\ \bibinfo {author} {\bibfnamefont {D.}~\bibnamefont {Gu\'{e}ry-Odelin}},\ }\bibfield  {title} {\bibinfo {title} {Strong saturation absorption imaging of dense clouds of ultracold atoms},\ }\href {https://doi.org/10.1364/OL.32.003143} {\bibfield  {journal} {\bibinfo  {journal} {Opt. Lett.}\ }\textbf {\bibinfo {volume} {32}},\ \bibinfo {pages} {3143} (\bibinfo {year} {2007})}\BibitemShut {NoStop}%
\end{thebibliography}%

\clearpage

\supplementarysection
\subsection{Sample purification}

After each sweep across the Feshbach resonance, the majority of the trapped atoms remain in their unbound state and are free to collide with the molecules, limiting their lifetime to less than 1 ms. This is why we need to remove the unbound atoms as quickly as possible after the resonance is crossed, and throughout this work we use two different methods to achieve this. During experiments presented in Figs. 2--4, after crossing the resonance, a microwave sweep was applied in order to transfer free Cs atoms from the $(f,m_f) = (3,3)$ to the (4,4) state. The sample was then subjected to a magnetic-field gradient of about 60 G/cm, which removed all the free K atoms, as well as all the Cs atoms in (4,4), which are anti-levitated under these conditions. However, the transfer to (4,4) in our setup has never been 100\% efficient, which resulted in up to 10,000 Cs atoms in (3,3) being left in the trap: a gradient of 60 G/cm over-levitates them, but not strongly enough. This number is sufficiently small to allow for molecular lifetimes on the order of 100 ms, which is enough for most of the measurements presented here. For the lifetime measurement (Fig.\ \hyperref[fig5]{5}), however, it was crucial to exclude as many loss sources as possible. Therefore, during those measurements we used an alternative method for removing Cs atoms: we first set the magnetic-field gradient close to the levitation gradient of the Feshbach molecules (45 G/cm), and then we reduced the depth of the optical trap until we stopped detecting any Cs atoms. The disadvantage of this method is that it also reduces the number of molecules remaining in the trap.

\begin{figure}
\includegraphics[width=0.48\textwidth,keepaspectratio]{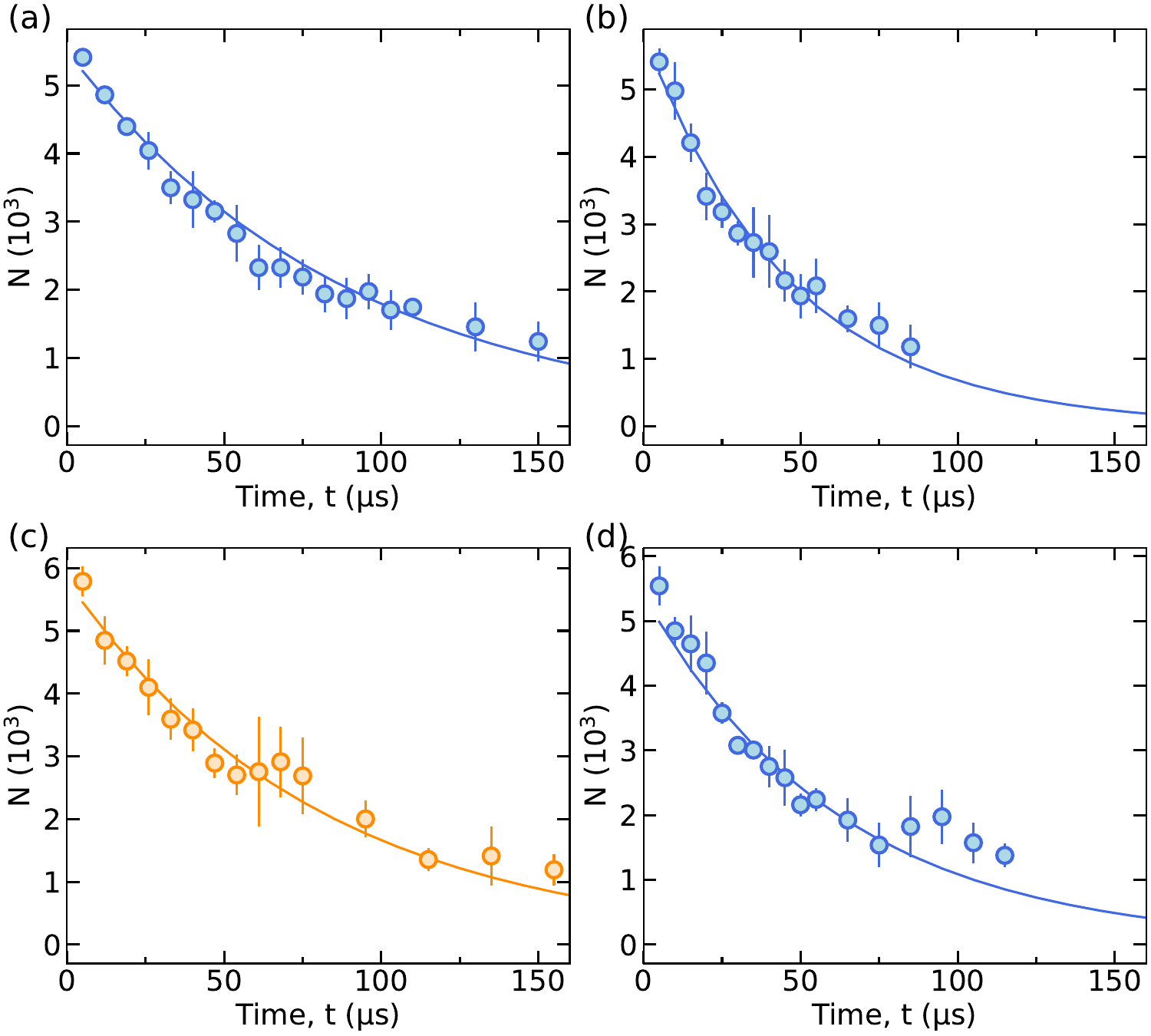}%
\caption{\label{figS1} Lifetimes of resonantly excited molecules. Number of molecules $N$ as a function of exposure time to a pump laser beam resonant with the $M_{J'} = -1$ (a), $M_{J'} = 0$ (b) or $M_{J'} = 1$ (c, d) component of the $A^1\Sigma_0-b^3\Pi_\Omega, v'=75, J'=1$ state. Blue (orange) points show measurements taken with the beam polarized horizontally (vertically). Solid lines are fits to Eq.\ 1 assuming zero detuning of the pump laser. The error bars reflect the standard deviation from five experimental runs.}
\end{figure}

\subsection{Excited state lifetimes}

The Rabi frequencies of the pump laser during addressing each of the hyperfine levels of $A^1\Sigma_0-b^3\Pi_\Omega, v'=75, J'=1$ presented in Fig.\ \hyperref[fig2]{2(b)} are calculated according to Eq.\ 1 using the natural linewidths extracted from the data in Fig.\ \hyperref[fig2]{2(b)} and the lifetimes of molecules exposed to the beam on resonance. The data used to extract these lifetimes for states with $M_{J'} = -1$ (a), $M_{J'} = 0$ (b) and $M_{J'} = 1$ (c, d) is shown in Fig.\ \hyperref[figS1]{S1}. As before, blue (orange) points show measurements taken with the beam polarized horizontally (vertically).

\subsection{Ground-state hyperfine structure}

The uncertainty of the rotational constant is estimated by numerically diagonalizing the hyperfine Hamiltonian at $|\vec{B}|$ = 343 G using the calculated parameters for KCs \cite{Aldegunde2017} and setting the unknown value of $g_r$ to 0.02, which is a typical value for bialkali molecules \cite{Aldegunde2008} and it is big enough to lift the degeneracy, but much smaller than the other energy splittings. The calculated energies of the states with $J''=0$ and $M''=2$ or $M''=4$ (accessible from the excited state with a vertically polarized beam) span a range of 506 kHz, and the energies of the states with $J''=2$ and $M''=3$ (accessible from the excited state with a vertically polarized beam) span a range of 544 kHz from the lowest state to the third highest state (which is where the leftmost peak in Fig.\ \hyperref[fig3]{3(b)} must be, since there are at least two states above it). This gives a total uncertainty of 1.05 MHz in the rotational splitting between $J''=0$ and $J''=2$, which is equal to $6B_0$.
 
\subsection{EIT modeling}

The model that we use to fit the EIT spectrum takes into account four levels: the initial, Feshbach state ($i$), the excited state ($e$), the rovibrational ground state ($g$) and a state representing the molecules in all other possible states, which we treat as lost ($0$). The dynamics of the molecules are then approximated by the following master equation \cite{Fleischhauer2005}: 
    
\begin{equation*}
\begin{split}
\dot{\rho} = \frac{i}{\hbar} [H,\rho] + \Gamma\left(\hat{\sigma}_{e0}\rho\hat{\sigma}^{\dagger}_{e0}-\frac{1}{2}\hat{\sigma}^{\dagger}_{e0}\hat{\sigma}_{e0}\rho - \frac{1}{2}\rho\hat{\sigma}^{\dagger}_{e0}\hat{\sigma}_{e0}\right) \\
+ \Gamma_{\textrm{eff}}\left(\hat{\sigma}_{gg}\rho\hat{\sigma}^{\dagger}_{gg}-\frac{1}{2}\hat{\sigma}_{gg}\rho - \frac{1}{2}\rho\hat{\sigma}_{gg}\right),
\end{split}
\end{equation*}

where $\rho$ is the density matrix, $\Gamma$ is the excited state linewidth $\hat{\sigma}_{e0} = \ket{0}\bra{e}$ is a transition operator from the excited state to the lost state, $\Gamma_{\textrm{eff}}$ is the effective decoherence rate between the initial and ground states and $\hat{\sigma}_{gg}$ is a projection operator onto the ground state. $H$ is the standard three-level Hamiltonian, which in the basis ($\ket{i}$, $\ket{e}$, $\ket{g}$), after applying the rotating-wave approximation, has the following matrix form:
\begin{equation*}
H = \frac{\hbar}{2}\left(\begin{array}{ccc} 0 & \Omega_p &  0 \\ \Omega_p & 2\Delta_p & \Omega_S \\ 0 & \Omega_S &  2\delta\end{array}\right),
\end{equation*}
where $\Omega_p$ and $\Omega_S$ are the Rabi frequencies of the pump and Stokes lasers, respectively, $\Delta_p$ and $\Delta_S$ are the detunings of those lasers from the transitions they are addressing, and $\delta=\Delta_p -\Delta_S$ is the two-photon detuning.

For the fit presented in Fig.\ \hyperref[fig3]{3(c)} we fix the excited state linewidth and the pump laser Rabi frequency at the values given by the previous measurements. For the fit in the inset we also fix the Stokes-laser Rabi frequency to the value extracted from the fit to the EIT spectrum and we assume perfect one- and two-photon resonance.

\subsection{Stokes-laser-induced Stark shifts}

\begin{figure}
\includegraphics[width=0.48\textwidth,keepaspectratio]{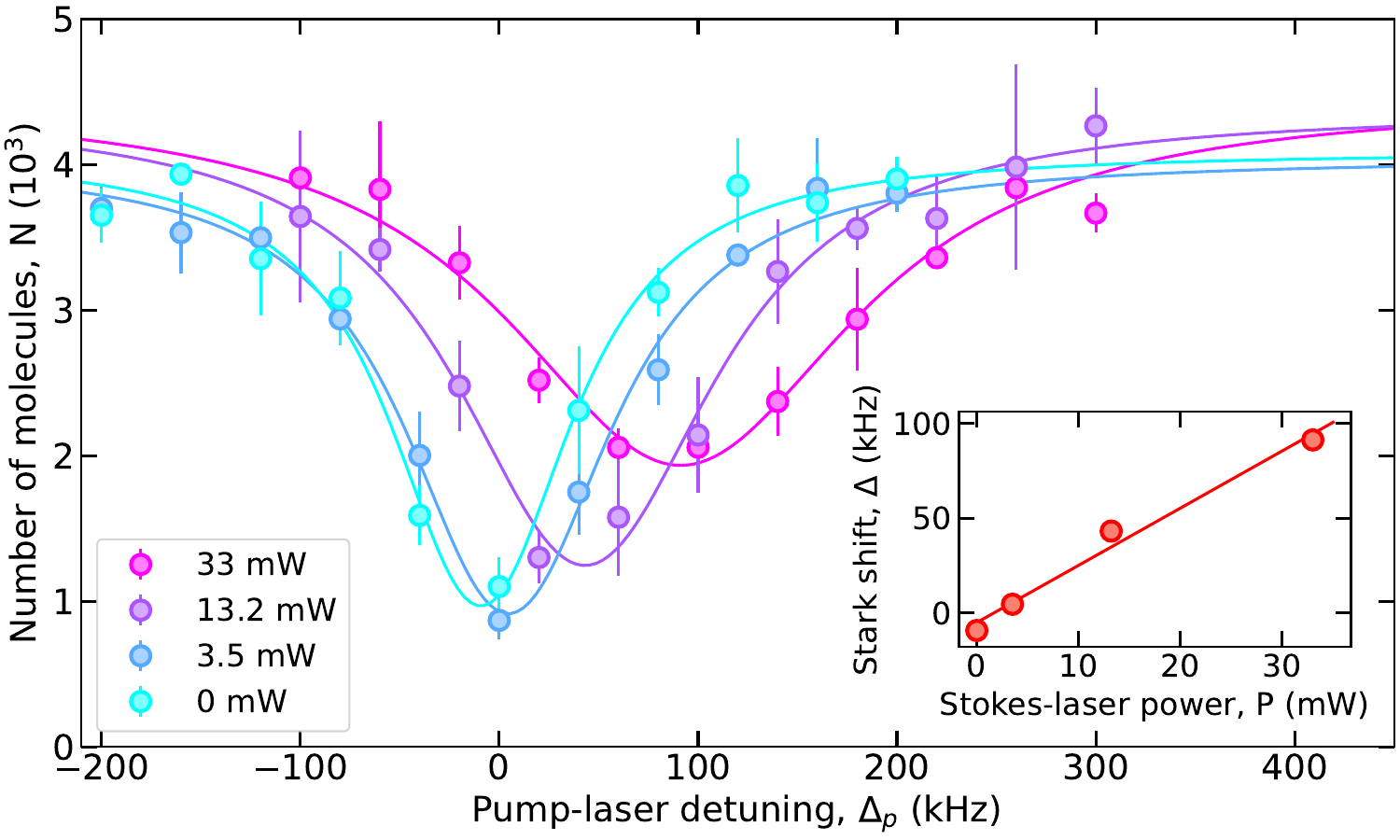}%
\caption{\label{figS2} Stark shift of the excited-state transition. Loss spectroscopy of the $A^1\Sigma_0-b^3\Pi_\Omega, v'=75, J'=1$ excited state at different optical powers of the Stokes beam detuned by 4 MHz from resonance with the ground state. Solid lines are fits to Eq.\ 1. The error bars reflect the standard deviation from three experimental runs. The inset shows the shift of the extracted resonance frequency as a function of the Stokes beam power. The solid line is a linear fit, and the error bars, given by the fits in the main figure, are smaller than the symbol size.}
\end{figure}

Due to the relatively low coupling of the excited state that we use for STIRAP ($A^1\Sigma_0-b^3\Pi_\Omega, v'=75, J'=1$) to the rovibrational ground state we need to shine the Stokes beam at peak intensities of up to 2 kW/cm$^2$ to achieve good transfer efficiencies. At the wavelength of 962.1 nm this results in a Stark shift of the atomic states of about 70 kHz for K and 200 kHz for Cs, and we expect the shift of the Feshbach state to be approximately the sum of these two. Given that the linewidth of our excited state is only about 80 kHz, and the intensity of the Stokes beam is ramped from its maximum value to zero during each STIRAP sequence, this shift could prevent us from satisfying the two-photon resonance condition and hence impact the transfer efficiency. In order to quantify the effect of the Stokes beam on the pump transition, we detune it from two-photon resonance by 4 MHz and shine it onto the sample with different powers, each time measuring the position of the excited state resonance. The results of these measurements are presented in Fig.\ \hyperref[figS2]{S2}. By fitting the same analytic model as in Fig.\ \hyperref[fig2]{2(b)} (Eq.\ 1) we extract the resonant frequencies at each value of the Stokes laser power and then fit these results to a linear function, which gives a slope of 3.0(3) kHz/mW and 48(7) Hz/(mW/cm$^2$). It is smaller by almost a factor of 3 than the expected shift of the Feshbach state, meaning that the excited state must be shifted in the same direction.

Remarkably, we have not observed any significant Stark shifts for the two-photon resonance, which implies that the AC polarizabilities of the Feshbach molecules and ground-state molecules must be very similar at 962 nm as well. Therefore, our STIRAP efficiency simulation includes a variation of the pump-laser detuning only.

\subsection{Trap frequencies in the ground state}

\begin{figure}
\includegraphics[width=0.48\textwidth,keepaspectratio]{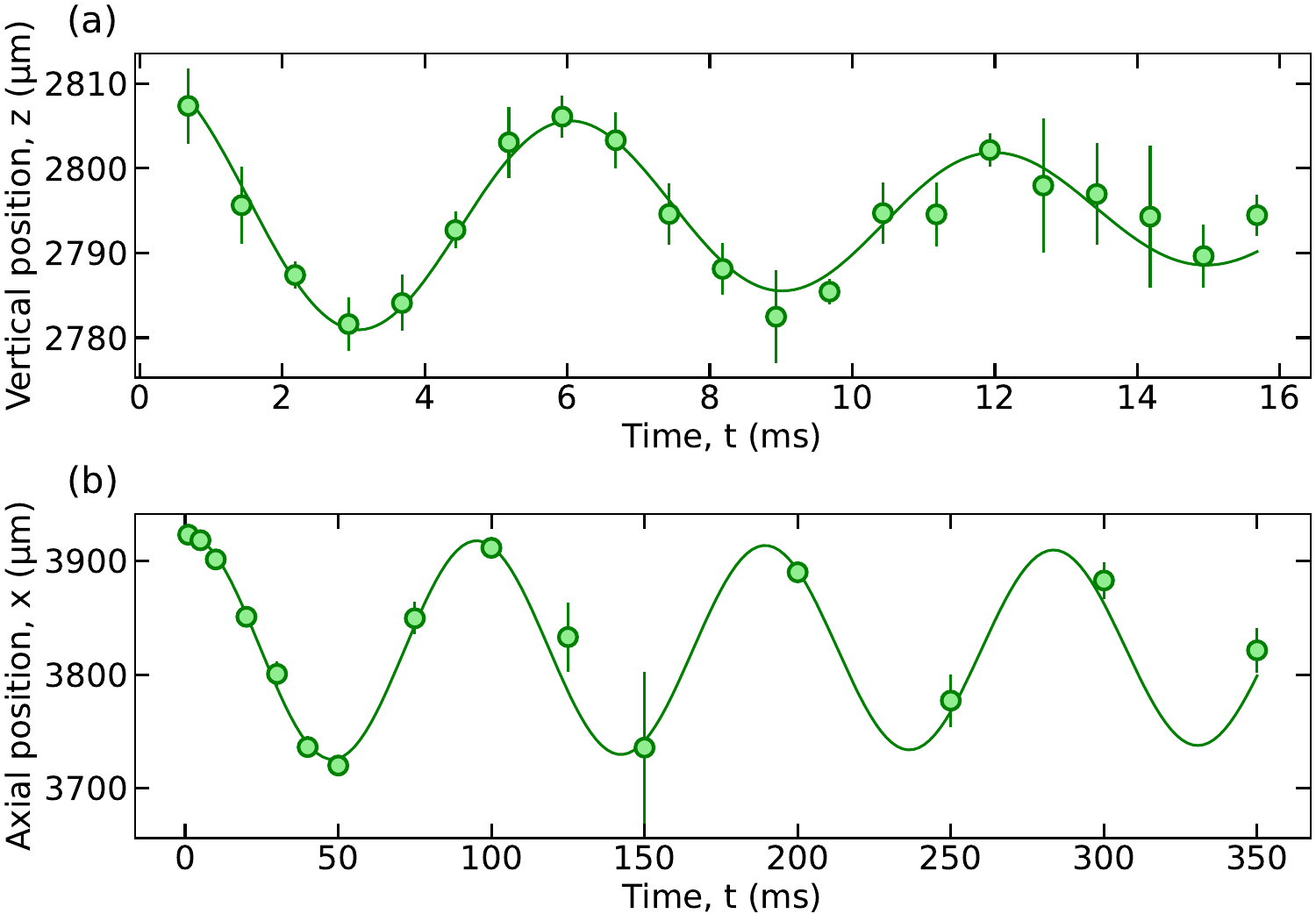}%
\caption{\label{figS3} Trap frequencies of the ground-state molecules. Position of the ground-state molecular cloud along the vertical direction (a) and the horizontal (axial) direction (b) as a function of hold time in the crossed trap. For the measurement of the vertical frequency (a) the cloud was first kicked using a third 1064-nm beam. For the axial frequency measurement (b) we used the same data as for Fig.\ \hyperref[fig5]{5}, since the cloud was already oscillating on its own. The solid lines are fits to damped sinusoidal functions, yielding frequencies of 168(4) Hz and 10.6(1) Hz, respectively. The error bars reflect the standard deviation from five experimental runs.}
\end{figure}

\begin{figure}
\includegraphics[width=0.48\textwidth,keepaspectratio]{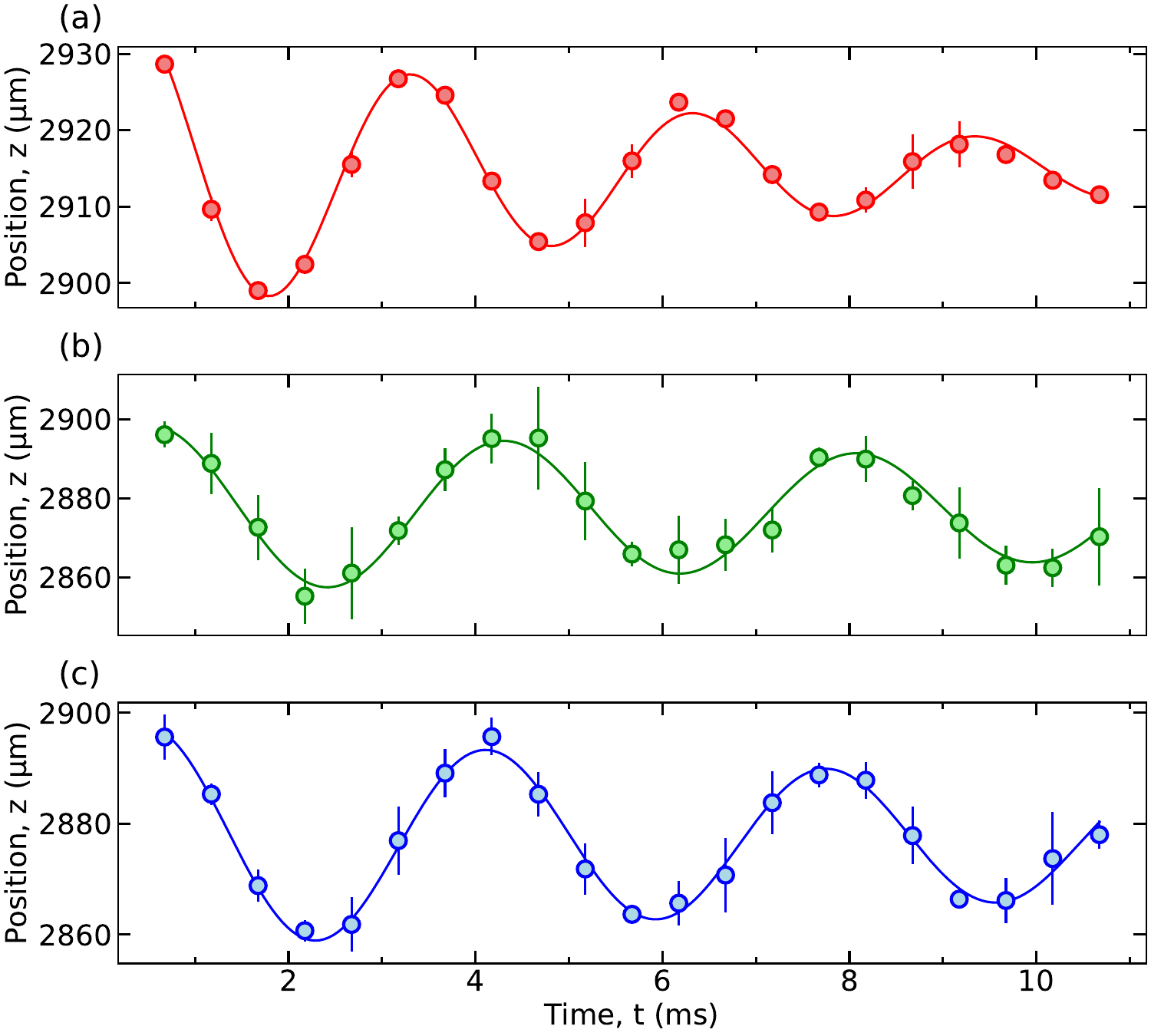}%
\caption{\label{figS4} Trap-frequency comparison. Vertical position of a cloud of potassium atoms (a), ground-state molecules (b) and Feshbach molecules (c) as a function of hold time after a vertical optical kick in a single-beam trap. The solid lines are fits to damped sinusoidal functions, yielding frequencies of 331(2) Hz, 265(3) Hz and 275(2) Hz. The error bars reflect the standard deviation from five experimental runs.}
\end{figure}

For most of the experiments presented here (Figs. 2--4) the molecules were trapped in a single-beam optical trap, in which the axial trapping was dominated by the residual magnetic-field curvature in our experimental chamber. However, once we transfer the molecules to the rovibrational ground state, their magnetic moment decreases almost to zero, which means that the axial trapping frequency is be determined by the optical intensity gradient along the beam and hence very low (around 1.5 Hz). Such weak trapping allows the molecules to spread along the beam on the timescales required to measure their lifetime, which eventually prevents accurate number measurements. This is why during experiments presented in Fig.\ \hyperref[fig5]{5} we crossed the original trapping beam, set to a power of 30(2) mW, with a perpendicularly propagating beam of a waist around 500 µm and 4 W of power. The vertical trap frequency of the ground-state molecules in this crossed trap was measured by kicking the cloud with a third 1064-nm beam and measuring its vertical position as a function of hold time. In the axial direction, the cloud oscillates on its own even before STIRAP, mainly due to sudden changes of $|\vec{B}|$ earlier in the sequence (we do not expect the momentum kick from the STIRAP beams to contribute significantly, as it corresponds to only $k_\textrm{B} \times 9$ nK of energy). Therefore, in order to measure the axial trap frequency, we directly extract the horizontal positions of the cloud from the data taken for Fig.\ \hyperref[fig5]{5}. The results of these measurements fit to damped sinusoidal functions with frequencies of 168(4) Hz and 10.6(1) Hz and are presented in Fig.\ \hyperref[figS3]{S3}.

In order to compare the optical polarizabilities of the ground-state and Feshbach molecules we also performed a radial trap-frequency measurement in a single-beam trap. Fig.\ \hyperref[figS4]{S4} shows the results of this measurement --- cloud position as a function of time after a kick with a second 1064-nm beam --- for (a) potassium atoms, (b) ground-state molecules and (c) Feshbach molecules, together with damped sinusoidal fits (solid lines) yielding frequencies of 331(2) Hz, 265(3) Hz and 275(2) Hz. The beam power was measured at 60(3) mW, so based on the potassium data we can calculate its waist to be 31.4(5) µm, which is consistent with the measurement in Fig.\ \hyperref[figS3]{S3} if gravity is taken into account. The polarizability of the ground-state molecules at 1064 nm is then calculated to be 78(11) Hz/(W/cm$^2$) and its ratio to the polarizability of the Feshbach molecules is 0.93(4).

\subsection{Compensation of the magnetic-field drifts}

\begin{figure*}
\includegraphics[width=\textwidth,keepaspectratio]{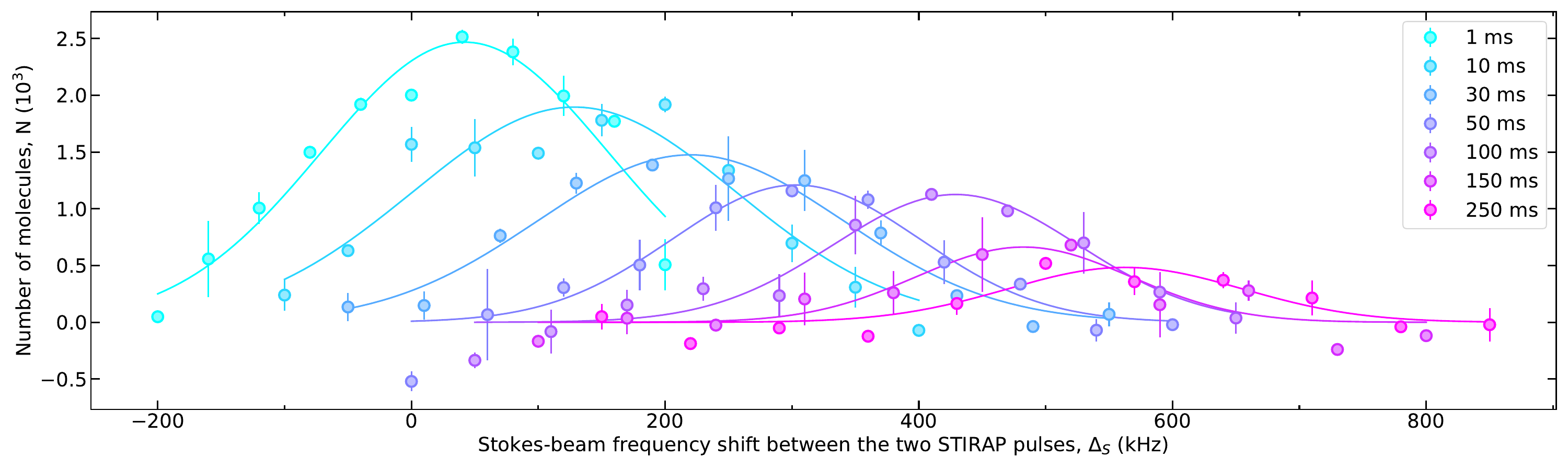}%
\caption{\label{figS5} Magnetic-field drift during the molecular lifetime measurement. Detected molecule number as a function of the shift between the RF frequencies sent into the Stokes laser switch AOM during forward and reverse STIRAP for different values of pulse separations. The solid lines are Gaussian fits with the background fixed at zero. The error bars reflect the standard deviation from two experimental runs.}
\end{figure*}

The response time of the magnetic field inside our experimental chamber to the changes in the current flowing through the coils is strongly limited by the eddy currents created in the steel walls of the chamber. In particular, after we sweep the magnetic field across the Feshbach resonance that we use for association, we observe a drift in the field even several hundreds of milliseconds later. This drift is sufficient to shift the STIRAP resonances by more than the two-photon linewidth determined by the pump and Stokes Rabi frequencies that we typically use. To compensate for this effect during the lifetime measurement, we change the RF frequency sent into the single-pass switch AOM of the Stokes beam between the forward and reverse STIRAP. The shift is small enough that it does not significantly reduce the diffraction efficiency of the AOM. Fig.\ \hyperref[figS5]{S5} shows the number of molecules detected after the round-trip STIRAP with different separations between the forward and reverse transfer as a function of the RF frequency sent to the AOM during the reverse transfer. The data in Fig.\ \hyperref[figS5]{S5} was used to determine the RF frequency to be sent to the AOM during the reverse STIRAP while taking the corresponding data points in Fig.\ \hyperref[fig5]{5} (the data used to determine this frequency for the other points is not shown for clarity).

\subsection{Loss rate measurement}
The two-body loss rate is calculated by fitting the data in Fig.\ \hyperref[fig5]{5} to a decay curve of the form $N(\tau) = N_0/(1+ \Gamma_2 \bar{n_0} \tau)$, where $\Gamma_2$ is the loss rate, $N_0$ is the initial number and $\bar{n_0}$ is the initial mean density.

The extracted value of the loss rate will inevitably be affected by multiple error sources. To quantify their effect, we split the quoted uncertainties into statistical (stat) and systematic (sys). The statistical uncertainty comprises fluctuations of the initial number (first point in the plot) of around 16\% and a 20\% error of the fitted $\Gamma_2$. The systematic error includes a 10\% uncertainty of the atom number calibration, which we estimate using the technique from \cite{Reinaudi:07}, as well as a fluctuation of the background atom count, amounting to $\pm8\%$ of the total number. It also includes a relative uncertainty of 14\% of the measured STIRAP efficiency, and the comparatively small uncertainties of the trap frequencies and temperature quoted in the main text.

\end{document}